\documentclass[aps,amsmath,amssymb, notitlepage, 
twocolumn,
nofootinbib,
]{revtex4-1}

\usepackage[pdftex]{graphicx}
\usepackage{dcolumn}
\usepackage{bm}
\usepackage{epsfig}
\usepackage{latexsym}
\usepackage{amsmath}
\usepackage{amsfonts}
\usepackage{amssymb}
\usepackage{color}
\usepackage{array}
\usepackage{framed}
\usepackage{esvect}
\usepackage{caption}
\usepackage{subcaption}
\usepackage{slashed}

\renewcommand{\v}[1]{\mathbf{#1}} 

\newcommand{\T}{\mathcal{T}}
\newcommand{\I}{\mathcal{I}}

\newcommand{\be}{\begin{equation}}
\newcommand{\ee}{\end{equation}}
\newcommand{\bea}{\begin{eqnarray}}
\newcommand{\eea}{\end{eqnarray}}

\newcommand{\ra}{\rangle}

\newcommand{\lp}{\left(}
\newcommand{\rp}{\right)}

\newcommand{\Z}{\mathbb{Z}}
\newcommand{\C}{\mathcal{C}}

\renewcommand{\vec}[1]{{\bf #1}}
\renewcommand{\hat}[1]{{\widehat #1}}

\def\nn{\nonumber\\}

\begin{document}

\title{Time-reversal symmetric $U(1)$ quantum spin liquids}
\author{Chong Wang and T. Senthil}
\affiliation{Department of Physics, Massachusetts Institute of Technology,
Cambridge, MA 02139, USA}
\date{\today}
\begin{abstract}
We study possible quantum $U(1)$ spin liquids in three dimensions with time-reversal symmetry. We find a total of 7 families of such $U(1)$ spin liquids, distinguished by the properties of their emergent electric/magnetic charges. We show how these spin liquids are related to each other. Two of these classes admit nontrivial protected surface states which we describe.  We show how to access all of the 7 spin liquids through slave particle (parton) constructions.  We also provide intuitive loop gas descriptions of their ground state wave functions. One of these phases is the `topological Mott insulator' conventionally described as a topological insulator of an emergent fermionic `spinon'.  We show that this phase admits a remarkable dual description as a topological insulator of emergent fermionic magnetic monopoles.  This results in a new (possibly natural) surface phase for the topological Mott insulator and a new slave particle construction. We describe some of the  continuous quantum phase 
transitions between the different $U(1)$ spin liquids.  
Each of these seven families of states admits a finer distinction in terms of their surface properties which we determine by combining these spin liquids with symmetry protected topological phases. We discuss lessons for materials such as pyrochlore quantum spin ices which may harbor a $U(1)$ spin liquid. We suggest the topological Mott insulator as a possible ground state in some range of parameters for the quantum spin ice Hamiltonian.

\end{abstract}
\maketitle

\tableofcontents

\section{Introduction}

There has been much recent interest in quantum spin liquid phases of systems of interacting magnetic moments.   These phases are fascinating examples of ground states characterized by long range quantum entanglement: the corresponding wave functions cannot be smoothly deformed into a product state of local degrees of freedom. Other examples of such long range entangled states include the celebrated fractional quantum Hall states. The structure of the long range entanglement dictates the excitation structure of the phase in just the same way as the familiar long range order (associated with broken symmetry) does in conventional ordered phases.  In particular a number of unusual excitations - for instance those with fractional quantum numbers or statistics, or gapless emergent gauge bosons - are possible in such phases.

This paper is concerned with a particular class of three dimensional quantum spin liquid that supports an emergent gapless `photon' as an excitation. 
It has long been recognized\cite{forster,wenbook} that such a  photon may be an emergent excitation of some underlying physical quantum many body system with short range interactions.  Specific microscopic models of quantum phases with emergent photons were constructed some time ago in Ref. \cite{bosfrc3d,wen03,hfb04,3ddmr,lesikts05,kdybk,shannon} in diverse systems. In addition to the photon these phases support other quasiparticle excitations which couple to the photon as  `electric' or `magnetic' charge. 

Interest in such phases has been revived following a recent  proposed experimental realization\cite{balentsqspice,gMFT}  in certain three dimensional pyrochlore oxides.  These are materials in which there are effective spin-$1/2$ degrees of freedom at the sites of the pyrochlore lattice. A class of such materials such as ${\rm Dy_2Ti_2O_7}$ or ${\rm Ho_2Ti_2O_7}$ have been studied extensively both in theory and experiment, and are adequately described within the framework of classical statistical mechanics\cite{gardnerrmp}.  Due to a combination of spin anisotropy and exchange interactions, the spins are constrained to satisfy an `ice rule' where on each tetrahedron of the pyrochlore lattice there are precisely two spins that point inward and two that point outward. The low energy physics takes place within the subspace of states satisfying this constraint. Hence these system have been dubbed `spin ice'. Quantum effects are known to be important\cite{gingrasrev} in a few such pyrochlore magnets which have 
hence been dubbed  `quantum spin ice'. Examples include ${\rm Yb_2Ti_2O_7}$, ${\rm Pr_2Zr_2O_7}$, and ${\rm Tb_2Ti_2O_7}$.  In particular in ${\rm Yb_2Ti_2O_7}$, the detailed microscopic Hamiltonian governing the interaction between the spins has been deduced through neutron scattering experiments\cite{balentsqspice,rossetal}.  The parameters of this Hamiltonian are such that quantum effects are surely present and will play a role in determining the low temperature physics.

It is well known\cite{balentsrev} that  {in the spin ice subspace the spins form oriented closed loops, and the subspace can be parametrized in terms of oriented loop configurations}. Classical spin ice systems are thus described as {\em thermally} fluctuating loop gases in three dimensions. The loops can be viewed as `magnetic' field lines of an artificial magnetic field. Defect configurations in the spin ice manifold such as a ``3-in 1-out" tetrahedron (where 3 spins point in instead of 2) correspond to end points of the loops and are then identified with magnetic monopoles\cite{castelnovo}. Such monopoles have been observed in experiments in the last few years\cite{morris,fennell}. 

In quantum spin ice materials it is natural to expect that the physics may be determined by {\em quantum} fluctuations of oriented loops. If these loops form a liquid phase where the loop line tension is zero the result is a quantum spin liquid. This spin liquid supports an emergent gapless photon. The associated magnetic field lines are simply the tensionless magnetic loops. Magnetic monopoles (the defect tetrahedra) are now gapped quasiparticle excitations where these field lines end.  Ref. \cite{balentsqspice} proposed that this physics may occur in ${\rm Yb_2Ti_2O_7}$.  

Knowledge of the precise microscopic Hamiltonian for ${\rm Yb_2Ti_2O_7}$ lends hope for a reliable theoretical assessment of this proposal and for quantitative comparisons to experiment. However the microscopic Hamiltonian is rather complicated and is hard to solve, either analytically or numerically. Further as we briefly review (see Appendix \ref{qspicemdl}) the parameters are such that it is not obvious that it is sufficient to just restrict to the spin ice manifold. 
Finally there is very little global symmetry in the model. The only good symmetries are time reversal and space group operations. 

What scope is there for theoretical progress in the absence of reliable methods to study the model Hamiltonian? One possibility is to deform the model to a limit where it's ground state may be reliably determined, say, by numerical methods. Approximate analytical methods can then be chosen to reproduce the known result in this limit.  They can then be extended to the realistic model with the hope that they capture the full phase diagram. 
For quantum spin ice such an approach has been pursued in Ref. \cite{gMFT} using a slave particle approach known as the `gauge mean field theory' (gMFT).  A reliable  limit is provided by considering the $XXZ$ spin-$1/2$ pyrochlore model in the Ising limit. This model can be studied through quantum Monte Carlo without a sign problem and the ground state is 
a $U(1)$ quantum spin liquid\cite{kdybk}. Further analytic arguments\cite{hfb04} strongly indicate the structure of the gapped $e$ and $m$ excitations.  The gMFT correctly reproduces this spin liquid ground state. Ref. \cite{gMFT} then extends 
this slave particle approach to obtain an answer for the full phase diagram of the model including the parameter regime determined in experiment. This mean field seems to show that the experimentally relevant parameters place the model in a conventional ferromagnetic state rather than a spin liquid. However this parameter regime is substantially different from the limit where gMFT is known to capture the correct ground state. It is hard to evaluate the accuracy of the gMFT prediction for the phase diagram away from this limit. In particular other slave particle mean field methods are available (for instance Schwinger bosons or fermions) and will lead to different phase diagrams. Further even when they lead to a $U(1)$ spin liquid it is not clear whether different slave particle methods lead to the same phase of matter.

In this paper, inspired by these developments,  we pose a different set of questions on which we are able to make solid progress. Rather than attempt to solve any particular microscopic model approximately we constrain the general properties of $U(1)$ quantum spin liquids\footnote{Here we only consider `spin liquids'  (or boson liquids) that can emerge in the Hilbert space of a purely spin (or boson) system.} in the presence of global symmetries and describe their physics. 
Specifically we focus on time reversal symmetric $U(1)$ quantum spin liquids, and determine the number of distinct phases and their properties. Time reversal is a robust physical symmetry, and the only internal symmetry in the model describing $Yb_2Ti_2O_7$.  We ignore space group symmetry both because it simplifies the problem and because it is less robust (due to disorder). To further simplify the problem we restrict to $U(1)$ liquid phases where the only gapless excitation is the photon. In particular the magnetic charge (dubbed the $M$ particle) and the electric charge (the $E$ particle) are gapped.\footnote{We also implicitly assume that, apart from the deconfined $U(1)$ gauge field,  there is no other co-existing topological order or source of long range entanglement.}   {We show that there are twenty-two phases which fall into seven distinct families of $U(1)$ spin liquids. The seven families of $U(1)$ spin liquids are distinguished by their bulk excitation spectrum, which we tabulate in Table.~\ref{u1gauge1}. Different phases in each family are distinguished by their surface states, and one can construct one phase from another in the same family by combining with a class of phases called symmetry-protected topological states. In most parts of this paper except Sec.~\ref{addspt}, we focus mainly on the seven families of states, which have clear physical differences in the bulk. Therefore we will often use the term ``phase'' instead of ``family of phases'' when the context is clear. } Most of the existing microscopic models describe only one of these phases which is also the one accessed by the gauge mean field theory of Ref. \cite{gMFT}.  

We describe the physics of these seven families of states and their interrelationships in many complementary ways.  We show how each of the seven families of states may be accessed through slave particle constructions. In some cases we provide more than one slave particle construction for the same phase. We describe the structure of the distinct ground states in terms of distinctions in the loop wave functions.  This leads to many interesting insights and to predictions for future numerical calculations. We determine the properties of protected surface states that some of these phases have. Given these solid results on the possible time reversal symmetric $U(1)$ spin liquids and their properties we may ask about how to distinguish them in experiments, and about which ones are likely for a particular microscopic model. We describe some experimental signatures that can help identify which (if any) of these spin liquids is realized.  We also provide some guides for relating to microscopic models.   A summary of our key results is in 
Sec. \ref{summ} below. 

We emphasize that the distinction between these phases is entirely a consequence of the unbroken time reversal symmetry. If this symmetry were absent then it is possible to go smoothly between any two of these phases. The distinction comes from different possible implementation of time reversal symmetry.  

\begin{table}[htdp]
\begin{tabular}{|c|c|c|c|}
\hline
Phase & Electric particle & Magnetic particle  \\ \hline
$E_bM_b$ & Boson & Boson  \\ \hline
$E_{bT}M_b$ & Boson, ${\cal{T}}^2=-1$ & Boson  \\ \hline
$E_fM_b$ & Fermion & Boson  \\ \hline
$E_{fT}M_b$ & Fermion, ${\cal{T}}^2=-1$ & Boson  \\ \hline
$E_bM_f$ & Boson & Fermion \\ \hline
$E_{bT}M_f$ & Boson, ${\cal{T}}^2=-1$ & Fermion  \\ \hline
$(E_{fT}M_f)_{\theta}$ & Fermion, ${\cal T}^2 = -1$ & $(q_m = 2)$ Fermion  \\ \hline
\end{tabular}
\caption{ Families of $U(1)$ quantum liquids with time-reversal symmetry labeled by properties of the `pure' electric and magnetic charges. $q_e$ and $q_m$ denote electric and magnetic charges, respectively. For the electric particle $(q_e, q_m) = (1,0)$ and for the magnetic particle $(q_e, q_m) = (0,1)$ except for the last row where it is $(0,2)$. For the last phase $(E_{fT}M_f)_{\theta}$ there are more fundamental `dyonic'  particles which have  $(q_e, q_m) =  ( \pm 1/2, \pm 1)$,  and are  bosons.  Both the pure electric charge and the pure magnetic charge indicated in the table can be built up as composites of the dyons in this phase. }
\label{u1gauge1}
\end{table}%


Our analysis will be strongly informed by recent progress\cite{avts12,hmodl,metlitski,scott,fidkowski3d,3dfSPT,3dfSPT2,maxvortex} in the theory of interacting generalizations of three dimensional topological insulators/superconductors (see Ref.~\cite{tsarcmp} for a review of aspects directly pertinent to this paper). It is now recognized that the topological band insulators are special examples of a class of quantum states of matter known as Symmetry Protected Topological (SPT) phases\cite{atav13,1dsptclass,chencoho2011}. These states are seemingly conventional in the bulk - they are gapped and have no exotic excitations but nevertheless have non-trivial surface states that are protected by symmetry. But what role do they play in describing the quantum spin liquids of interest in this paper? The answer is that starting with one kind of $U(1)$ spin liquid we may generate others by putting one of the emergent quasiparticles ($E$ or $M$) into an SPT state. { For $U(1)$ quantum spin liquids this point of view was initiated in a previous paper\cite{hmodl} by the present authors. Ref. \cite{hmodl}  considered SPT states of bosonic particles, and demonstrated their utility in understanding some $U(1)$ quantum spin liquids. This point of view will be fully developed in the present paper    and lead to a complete and more insightful 
description of all time reversal invariant $U(1)$ spin liquids with gapped matter.  In particular we will exploit recent exciting developments on fermionic SPT states\cite{fidkowski3d,3dfSPT,3dfSPT2,maxvortex} that  were not understood when Ref. \cite{hmodl} was published to obtain this complete picture.}

It is important to point out that there is no one-to-one mapping between SPT phases with global $U(1)$ and time reversal symmetries, and $U(1)$ quantum spin liquids with time reversal. They both have different classifications. For example, we will show that two different SPT states reduce after gauging to the same physical $U(1)$ spin liquid.



\section{Summary of results}
\label{summ}
Here we briefly summarize some of our key results. This section will also serve as an outline for the rest of the paper. 

\begin{enumerate}
\item
We first establish that there are  $7$ distinct families of time reversal invariant $U(1)$ liquid phases in $3d$ distinguished by their bulk spectra in Sections \ref{prelim}, \ref{desc} and \ref{tmi}.  A partial characterization of these phases is obtained by plotting the spectrum of emergent quasiparticles - the charge-monopole lattice - in the $U(1)$ gauge theory. We show that six of these $7$ families have the charge-monopole lattice of 
Fig. \ref{cmlat1} while the remaining one has the charge-monopole lattice of Fig. \ref{cmlat2}.  We provide a first cut description of these $7$ families of phases in these sections and relate them to existing constructions of $U(1)$ liquids.  One of these phases (dubbed $E_bM_b$) is the state accessed by gMFT while some others are states accessed by Schwinger boson or Abrikosov fermion constructions. The unique family described by Fig. \ref{cmlat2} includes the so-called `Topological Mott Insulator' discussed in Ref. \cite{pesinlb}.   For reasons described later this is denoted $\left(E_{fT}M_f\right)_\theta$ in this paper. 
The family denoted $E_{bT}M_f$ has not been described explicitly in the literature. 

\item
We describe how these phases are related to each other in Sec. \ref{relation}.  This is enabled by recent advances in our understanding of SPT phases of bosons/fermions with global $U(1)$ and time reversal symmetries. We continue the point of view adopted in our previous work\cite{hmodl} showing that given one $U(1)$ liquid we can obtain others from it by putting one of the emergent quasiparticles in an SPT phase.  We are thus able to obtain a rather complete understanding of how these $7$ phases are related to each other. 

\item
The conventional description of the  $\left(E_{fT}M_f\right)_\theta$ (the `Topological Mott Insulator') is that it is a topological insulator formed by emergent fermionic Kramers doublet spinons that are coupled to the $U(1)$ gauge field as electric charge.  We show in Sec. \ref{relation} that this phase has a remarkable dual description as a topological insulator of emergent fermionic magnetic monopoles. 

\item
We discuss the possibility of protected surface states at the interface with the vacuum for these spin liquids in Sec. \ref{surf}. In subsection \ref{whysurface} we describe criteria that determine when such  protected surface states will form. We argue that precisely two of the $7$ families ($\left(E_{fT}M_f\right)_\theta$ and $E_{bT}M_f$) are required to have a non-trivial surface state.  The possibility of a surface spinon Dirac cone for the 
$\left(E_{fT}M_f\right)_\theta$ is well known\cite{pesinlb}. The dual description of this phase as a monopole topological insulator naturally leads to an alternate possible `dual' surface state where there are an odd number of gapless monopole Dirac cones (and no spinon Dirac cone). We then describe the surface of $E_{bT}M_f$ - the simplest possibility is to have two monopole Dirac cones. 

\item
For any  given bulk spectrum there can be more than one phase corresponding to distinct surface properties.  When these are taken into account we find a total of $22$ distinct phases. These are obtained from the $7$ basic phases by combining them with SPT phases of bosons/spins protected by time reversal alone. Interestingly, in some cases, the spin liquid can `absorb' an SPT phase so that the combination is not in a distinct phase. In other words not all $T$-reversal symmetric SPT phase stay distinct from trivial phases when combined with a spin liquid. Similar phenomenon also appears in two dimensional $\mathbb{Z}_2$ spin liquids\cite{maxprivate}.

\item
We show, in Sec. \ref{prtn} how to access all the $7$ basic phases through parton constructions on spin models. In particular we show how the standard fermionic parton construction of spin-$1/2$ systems enables access to $5$ of the $7$ phases (the exceptions being $E_bM_b$ described by gMFT and the $E_{bT}M_b$ described by Schwinger bosons). For the topological Mott insulator $\left(E_{fT}M_f\right)_\theta$ we describe a dual parton construction in terms of monopoles that is distinct from the conventional one in terms of spinons. With a view toward obtaining input on microscopics, we obtain some no-go results on these parton constructions in  Sec. \ref{nogo} if the physical system consists of Kramers doublet spin-$1/2$ degrees of freedom. 

\item
We provide an intuitive physical picture of the ground state wave function for these spin liquids in terms of fluctuating loop configurations of electric or magnetic field lines in Sec. \ref{loops}. 

\item
We describe some of the remarkable continuous quantum phase transitions between these different spin liquids. We particularly focus on phase transitions out of the topological Mott insulator $\left(E_{fT}M_f\right)_\theta$.  We provide a theory for a second order transition from this phase to others where the electric charge is a boson (either Kramers singlet or doublet). We also provide a theory for a different second order phase transition between two different phases where the electric charge changes from Kramers doublet to Kramers singlet. 

\item
In Sec. \ref{mat} we consider the relevance of these results to current and future realizations of $U(1)$ spin liquids in experimental systems. For pyrochlore spin ices based on Kramers doublet spin systems we discuss the possible $U(1)$ spin liquids that may obtain. Apart from the one suggested by gMFT, we argue, based on the parton construction, that the other natural candidate is the topological Mott insulator $\left(E_{fT}M_f\right)_\theta$.  A strong coupling expansion of the lattice parton Hamiltonian coupled to the $U(1)$ gauge field yields at leading order a spin Hamiltonian of the form appropriate to the pyrochlore spin ices but with parameters different from the ones where gMFT is expected to be reliable. We suggest on this basis that some pyrochlore spin ices may be in the topological Mott insulator phase.  

We also describe some experimental distinctions between these phases which may be useful in identifying them. 

Several appendices contain peripheral details. 

\end{enumerate}

\section{Preliminaries}
\label{prelim}

We begin with some simple but powerful observations. We are interested in time reversal symmetric $U(1)$ liquids of spins/bosons in which the only gapless excitation is the photon. To distinguish different phases it is appropriate to focus on the gapped emergent quasiparticles that couple as electric or magnetic charges to the photon. Time reversal symmetry constrains the possibilities in many important ways as we now describe. 

\subsection{Charge-monopole lattice}
\label{lattice}

We denote the electric charge $q_e$ and magnetic charge $q_m$.  We use notation in which  the total electric flux is $4\pi q_e$ and the total magnetic flux is $2\pi q_m$.  
To be general we must allow for the most fundamental emergent particles to be ``dyons", {i.e} particles that carry both electric charge and magnetic charge.   
For any pair of dyons with charges $(q_e = Q_e, q_m= Q_m)$ and $(q_e = Q_e', q_m = Q_m')$ there is a generalized Dirac quantization condition\cite{zwanziger,schwinger}: 
\begin{equation}
\label{diracq}
Q_e Q_m' - Q_m Q_e' = n
\end{equation}
where $n$ is an integer. 

For each particle with charges $(Q_e, Q_m)$ there will be an antiparticle with charges $(-Q_e, - Q_m)$. Note that the particle and antiparticle automatically satisfy the Dirac quantization condition. 
We will use the natural convention that under  time reversal the magnetic fields are odd and the electric fields are even. Then for any particle with charges $(Q_e, Q_m)$ there is a time reversed partner with charges $(Q_e, - Q_m)$. 
Applying the Dirac quantization condition to these two particles we obtain the restriction
\begin{equation}
\label{dyonq}
2Q_e Q_m = \text{integer}
\end{equation}

By combining $(Q_e, Q_m)$ with $(Q_e, -Q_m)$ we can produce a particle that is a pure electric charge $(2Q_e, 0)$. Similarly by combining $(Q_e, Q_m)$ with $(-Q_e, Q_m)$ (the antiparticle of the time reversed partner) we obtain a pure magnetic charge $(0, 2Q_m)$.  
Thus time reversal invariance guarantees that there are always both pure electric and pure magnetic charges in the theory. 

Consider the smallest pure electric charge.  We choose units in which this has $q_e = 1$ (and by definition has $q_m = 0$).   Let the smallest pure magnetic charge have strength $g$ (and $q_e = 0$).  Applying the Dirac condition to the pure electric charge and the pure magnetic charge we get 
\begin{equation}
\label{nondyong}
g = \text{integer}
\end{equation}
If there are no other restrictions the smallest allowed $g$ is $1$.  As is well known the Dirac condition requires that pure electric and charges are quantized to be integers (in our units). If there are dyons with charges 
$(Q_e, Q_m)$ it follows that $2Q_e$ must be an integer.  Thus we have two basic possibilities $Q_e = 1$ or $Q_e = \frac{1}{2}$ . In the former case there are no further restrictions on $g$ beyond Eq.~\eqref{nondyong} and we have $g = 1$. In the latter case we can apply Dirac quantization to the $(0,g)$ and $(\frac{1}{2}, Q_m)$ particles to obtain 
\begin{equation}
g = 2\times\text{integer}
\end{equation}
Thus if there are charge-$1/2$ dyons the minimum pure magnetic charge is $2$. Further we must have $Q_m = 1$ for the charge-$1/2$ dyon.

\begin{figure}
\begin{center}
\includegraphics[width=1.9in]{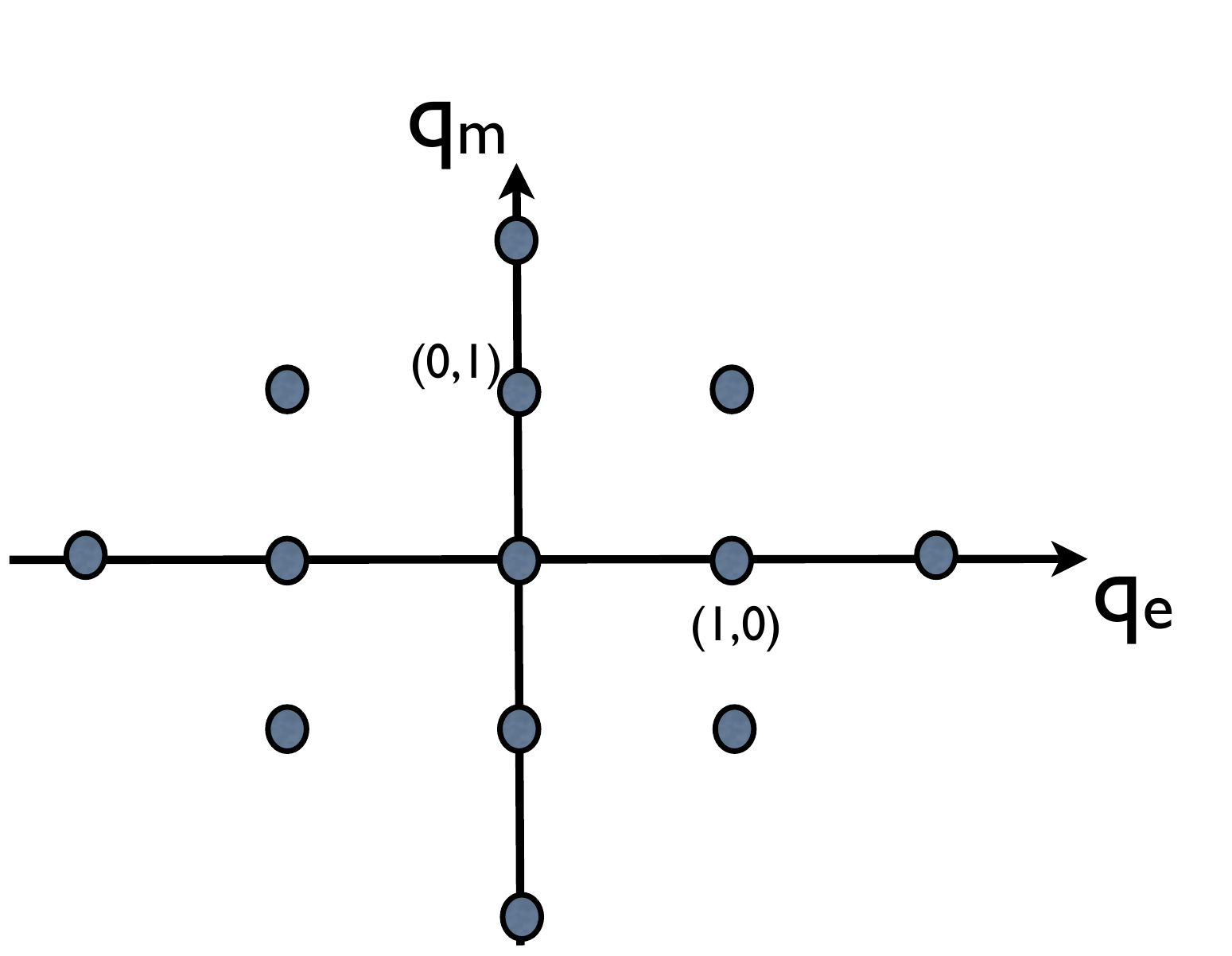}
\end{center}

\caption{Charge-monopole lattice at $\theta = n\pi$ with $n$ even.  }
\label{cmlat1}

\end{figure}

\begin{figure}
\begin{center}
\includegraphics[width=1.9in]{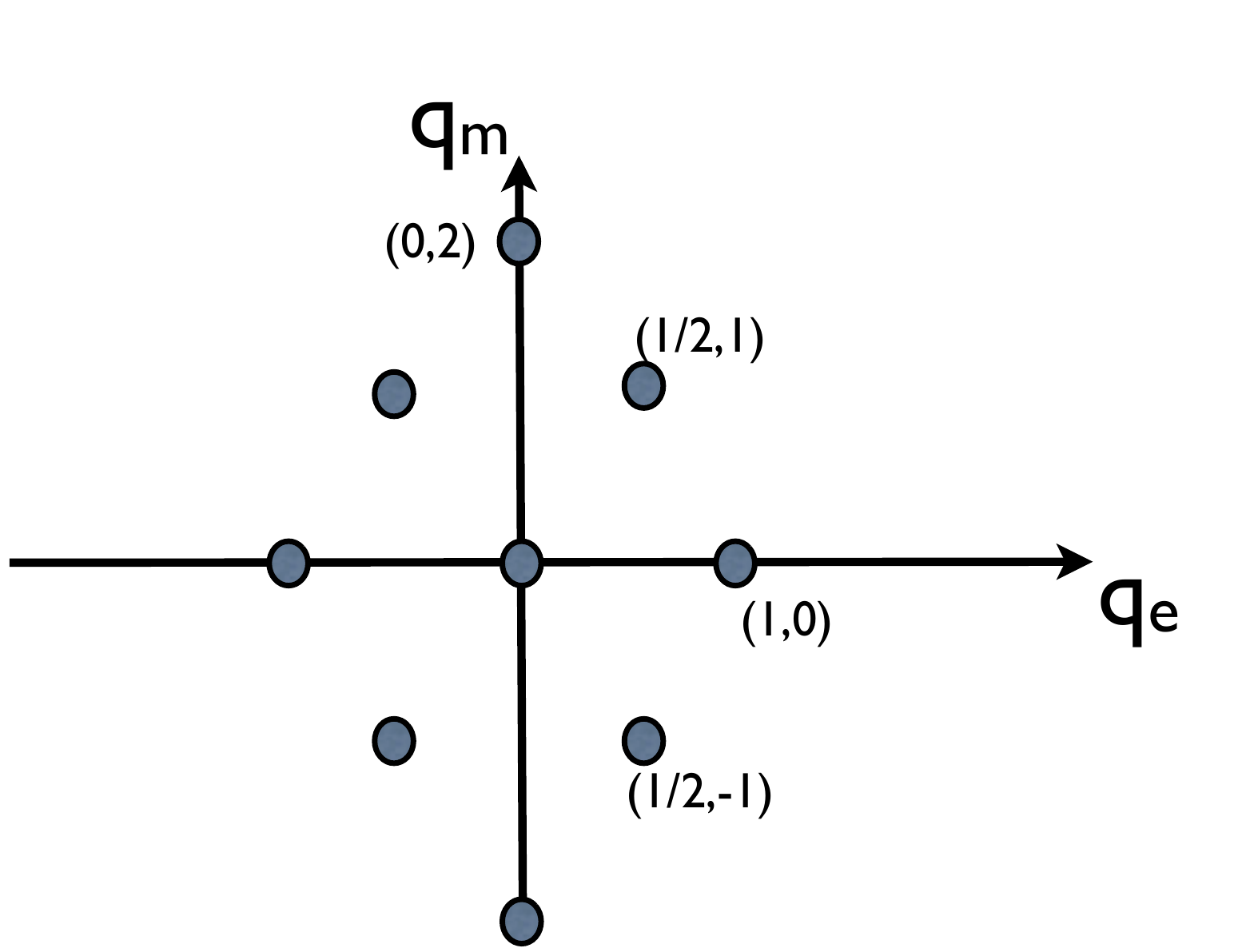}
\end{center}

\caption{ Charge-monopole lattice at $\theta = n\pi$ with $n$ odd. }
\label{cmlat2}

\end{figure}

  We thus have two classes of possible states which are distinguished by the geometry of the lattice of allowed charges and monopoles. In one class the charge-monopole lattice is as shown in Fig. \ref{cmlat1}. Here all emergent quasiparticle excitations are obtained from two elementary quasiparticles - the $E$ particle with $(q_e, q_m) = (1,0)$ and the $M$ particle with $(q_e, q_m) = (0,1)$. In the second class the charge-monopole lattice is shown in Fig. \ref{cmlat2}. Here the full set of emergent particles can be built out of the two dyons with $(q_e, q_m) = (\pm \frac{1}{2}, 1)$.

These two general possibilities can also formally be distinguished in terms of the low energy effective Lagrangian for the photon after integrating out the $E$ and $M$ particles.  This takes the form
\begin{equation}
{\cal L}_{eff} = {\cal L}_{Max} + {\cal L}_\theta
\end{equation}
The first term is the usual Maxwell term and the second is the `theta' term:
\begin{equation}
{\cal L}_\theta = \frac{\theta}{4\pi^2} \v{E}\cdot\v{B}
\end{equation}
where $\v{E}$ and $\v{B}$ are the electric and magnetic fields respectively. 

As is well known time reversal restricts the allowed values to $\theta = n\pi$ with $n$ an integer.  
$n$ even corresponds to Fig. \ref{cmlat1} and $n$ odd to Fig. \ref{cmlat2}\cite{witten}. 

Each of these two charge-monopole lattices can potentially be realized in several ways depending on the statistics of the quasiparticles and transformation under time reversal. We next describe the constraints on these. 

\subsection{Quasiparticle statistics and symmetry realization}

For a spin or boson system, in the {\em microscopic} Hilbert space, excitations created by local physical operators must clearly be bosonic.  The emergent quasiparticles $E$ or $M$ are however not created by local operators. For instance to create an $E$ particle it is also necessary to create the `electric' field lines that emanate from it and which extend out to arbitrary long distance. A  formal way of describing this is to say that 
the `creation' operator for $E$ (or $M$) alone is not gauge invariant. Creating $E$ without the associated electric field violates Gauss law and hence is not in the physical Hilbert space. As $E$ and $M$ are not created by local physical operators there is no restriction that they must be bosonic. In three space dimensions they can thus be either bosons or fermions.  Very recently it has been shown\cite{3dfSPT,mcgreevy} however that in a strictly $3d$ spin/bosonic system (as opposed to systems that can only appear the boundary of a $4+1$ dimensional system) that $E$ and $M$ cannot simultaneously be fermionic. We will return to this below. 

Time reversal  symmetry acts in a simple way on physical states in the Hilbert space of spins/bosons. The time reversal operator ($T$) is anti-unitary and satisfies $T^2 = +1$ on all physical states. This should be contrasted with electronic systems where $T^2 = -1$ for an odd number of electrons which leads to Kramers degeneracy. Let us now discuss the possible action of time reversal on the emergent $E$ and $M$ particles.  Quite generally the structure of the emergent Maxwell equations implies that the electric charge is even while the magnetic charge is odd under time reversal\footnote{Strictly speaking what we call electric and what we call magnetic is a matter of convention: the $U(1)$ gauge theory is self-dual so that we can interchange the definition of $E$ and $M$. Maxwell's equations imply that the electric and magnetic charge transform oppositely under time reversal. It is natural to adopt the convention that the magnetic charge is time reversal odd.}.  Thus the $E$ particle and its time reversed 
partner $TE$ only differ by a local operator. Then $T^2$ acting on the $E$ particle has a well defined value. Now as $E$ itself is not local, it could have $T^2 = -1$ and hence be a Kramers doublet (we review some more details in Appendix~\ref{tre}). In contrast $M$ and $TM$ do not differ by a local operator. Then there is no meaning to asking whether $M$ is Kramers or not. Specifically  $T^2$ acting on $M$ can be shifted by a gauge transformation to have any value\cite{hmodl}. 

Finally though $E$ or $M$ may be a fermion, and $E$ may be a Kramers doublet, composite excitations formed out of them that carry zero electric and magnetic charge are physical excitations, and hence must be bosonic Kramers singlets.

Starting with these simple but powerful observations we proceed to describe all the distinct time reversal invariant $U(1)$ spin liquids where the photon is the only gapless excitation.

\section{Phases with $\theta = 0$}
\label{desc}
We first describe phases in which the parameter $\theta = 0$, {\em i.e} the charge-monopole lattice is given by Fig. \ref{cmlat1}. Here 
we distinguish two broad classes of phases depending on whether the $M$ particle is a boson or fermion. We describe each in turn. 

\subsection{Bosonic monopole} 
\label{bosonM}
It is clear first that there are $4$ distinct phases where $M$ is a boson. The $E$ particle may either be a boson or a fermion and be Kramers singlet or doublet.  
Let us  understand better these $4$ phases. We label them  $E_bM_b,E_{bT}M_b,E_fM_b,E_{fT}M_b$, respectively, with the subscripts $b,f$ describing the statistics, and the symbol $T$ referring to Kramers degeneracy.  
Some of these are obtained through familiar constructions. 

The $E_bM_b$ phase is the one constructed in most of the existing microscopic models\cite{bosfrc3d,hfb04,3ddmr,lesikts05,kdybk,shannon}. It is also the state accessed by the gauge mean field theory of Ref.\cite{gMFT}.  If in addition to time reversal there is a global $U(1)$ symmetry, an intuitive way to understand this phase was described in Ref. \cite{lesikts05} by obtaining it from a proximate long range ordered phase ({\em i.e} with broken $U(1)$ symmetry) through proliferating appropriate vortex loops. 

The phases $E_{bT}M_b, E_{fT}M_b$ are accessed by the standard Schwinger boson or Abrikosov fermion representation of the physical spin. It is well known that in $3d$ these representations can lead to stable $U(1)$ spin liquid phases with gapped electric and magnetic charges.

For spin systems with spin rotation symmetry, it is instructive to obtain  $E_{bT}M_b$ by starting with a semiclassical description of a Neel antiferromagnet as follows. Consider a collinear Neel state of a quantum antiferromagnet in $3d$. The corresponding order parameter manifold is $S^2$. An effective field theory description of the long wavelength fluctuations of the Neel order parameter is provided by the quantum non-linear sigma model in $3+1$ space-time dimensions with the Euclidean action:
\begin{equation}
S_{NL\sigma M} = \frac{1}{2g} \int d^3x d\tau \left( (\vec \nabla \hat{n})^2 + \frac{1}{c^2} \left(\partial_\tau \hat{n} \right)^2 \right)
\end{equation}
Here $\hat{n}$ is the local orientation of the Neel vector, and $c$ is the spin wave velocity. 
In $3d$ the order parameter manifold allows for point defects known as `hedgehogs' corresponding to $\Pi_2(S^2) = Z$. This Neel state may be quantum disordered without proliferating these hedgehogs. 
A convenient framework to describe this is through a $CP^1$ representation: $\hat{n} = z^\dagger \vec \sigma z$ where $z$ is a two-component complex spinor. Importantly under time reversal $\hat{n} \rightarrow - \hat{n}$
and $z_\alpha \rightarrow i\sigma^y_{\alpha \beta} z_\beta$. Thus $z$ is a Kramers doublet. The $z$ representation introduces a $U(1)$ gauge redundancy ($z(x, \tau) \rightarrow e^{i\theta(x,\tau)}z(x,\tau)$). The sigma model action represented in terms of $z$ naturally includes a compact $U(1)$ gauge field $a_\mu$. It is well known that the monopoles of the $a_\mu$ correspond, in the Neel ordered state, to the hedgehogs of the $\hat{n}$ field. Quantum disordering the Neel state corresponds to gapping out the $z$-particles. If in addition the monopoles stay gapped the result is precisely a $U(1)$ spin liquid. Further the $z$ get identified with the $E$ particle and the $M$ with the remnants of the hedgehog. Clearly $E$ is a Kramers boson. In the semiclassical limit the hedgehog is also a boson and consequently so is the $M$ particle in the $U(1)$ spin liquid. Thus the phase we obtain is precisely the $E_{bT}M_b$ $U(1)$ spin liquid.

Finally, a microscopic model for the $E_fM_b$ phase was constructed in Ref.~\cite{levinu1f}.  In Sec.~\ref{relation} we describe how it is related to the other phases, in particular to the simple $E_bM_b$ phase. 

\subsection{Fermionic monopole}
\label{fermionM}

We now consider cases in which the $M$ particle carries fermion statistics.  
If $M$ is a fermion, recent work\cite{3dfSPT,mcgreevy} shows that the $E$ particle cannot also be a fermion in a strictly three dimensional system.  With a bosonic $E$ particle there are however still two distinct possibilities corresponding to whether it has $T^2 = +1$ or $T^2 = -1$, {\em i.e} whether it is a Kramers singlet or doublet. In obvious notation we label these two phases $E_bM_f$ and $E_{bT}M_f$.  

We show in Section~\ref{prtn} how to access these phases through a parton construction. 


\section{Phases with $\theta = \pi$: ``Topological Mott insulator"}
\label{tmi}
We now discuss  time reversal symmetric $U(1)$ spin liquids  with $\theta = \pi$.  We will see that there is precisely one such phase. 

First let us discuss the statistics of the elementary dyons with charges $(q_e, q_m) = ( \frac{1}{2}, \pm 1)$. We note that these are interchanged under time reversal. Thus they are required to have the same statistics, {\em i.e} they are both bosons or both fermions. However we can argue that they cannot both be fermions.  To see this most simply we note that  the  $((\frac{1}{2}, 1)$ and the $(\frac{1}{2}, -1)$ dyon are relative monopoles, {\em i.e} each one sees the other the way an electric  charge sees a monopole. If they were both fermionic we would have a realization of the ``all-fermion" $U(1)$ gauge theory in a strictly $3+1$ dimensional system which we know is not possible\cite{3dfSPT,mcgreevy}. Therefore we conclude that both these dyons must be bosons. 

Now consider the bound state of these two dyons.  As this has $q_e = 1, q_m = 0$ we identify it with the `elementary' pure electric charge in this phase. Precisely this bound state was analysed recently in Refs. \cite{3dfSPT,metlitski} while studying correlated topological insulators, and shown to be a fermion with $T^2 = -1$, {\em i.e}, a Kramers doublet. In brief these two dyons see each other as relative monopoles. This leads to the Fermi statistics of their bound state. The Kramers degeneracy can be simply understood by first calculating the angular momentum of the $U(1)$ gauge field. It is readily seen that this is quantized to be $1/2$. Combining this with the observation that time reversal inverts the relative coordinate of the two dyons leads to $T^2 = -1$ for their bound state.  Thus the statistics and symmetry properties of the elementary electric charge are uniquely determined for this charge-monopole lattice. 

Next consider the elementary pure magnetic charge which has $q_e = 0, q_m = 2$. This can be obtained as the bound state of the $((\frac{1}{2}, 1)$ and $((-\frac{1}{2}, 1)$ dyons. These are also relative monopoles and hence their bound state is a fermion. Now time reversal does not interchange these two dyons and hence the argument above for the Kramers structure of the pure electric charge does not apply. This is  of course in line with the earlier argument that it is meaningless to ask if $M$ particles are Kramers or not. 

We thus see that the structure of both the elementary electric charge and the elementary magnetic charge are uniquely determined for this charge-monopole lattice. In addition the statistics and symmetry properties of the elementary dyons is also fixed. Thus there is precisely one time reversal symmetric $U(1)$ spin liquid phase corresponding to $\theta = \pi$. 
Given these properties of the elementary pure electric and magnetic charges we denote this phase $\left(E_{fT}M_f \right)_\theta$. The subscript $\theta$ is a reminder that these pure charges are composites of more fundamental dyons.

This phase may be constructed within slave particle methods. 
Let us begin with $E_{fT}M_b$ where the $E$ particle is a 
Kramers doublet fermion, and has a conserved electric charge that is even under time reversal.  We then put this $E$ particle into a topological band insulator phase.  It is well known\cite{qi} that the topological band structure leads to a $\theta = \pi$ term in the action for a $U(1)$ gauge field that couples to the $E$ particle. This slave particle construction of the $\left(E_{fT}M_f \right)_\theta$ $U(1)$ spin liquid was discussed in Ref. \cite{pesinlb} and dubbed the `topological Mott insulator'. Ref. \cite{pesinlb} also suggested that this phase may be realized in $Y_2Ir_2O_7$ though this has turned out to be unlikely. 

Later we will describe a completely different slave particle construction of this phase. 



\section{Relationship between the phases}
\label{relation}
\begin{figure}
\begin{center}
\includegraphics[width=2.5in]{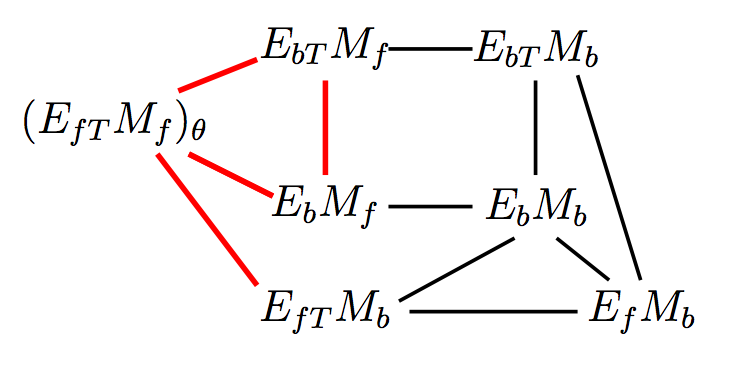}
\end{center}
\caption{Relationship between different $U(1)$ spin liquids. Two phases connected through a line share a common fundamental particle ($E$ or $M$), and can be viewed as different SPT phases formed by the common particle. In Sec.~\ref{transitions} we describe some intersting continuous phase transitions between the phases connected through thick red lines. }
\label{phaserelation}
\end{figure}

We have thus completed the description of Table \ref{u1gauge1}.   We now describe how these seven different phases are related to each other.  In our previous work (Ref. \cite{hmodl}) we addressed this for a subset of these phases, and showed that they can be related to different Symmetry Protected Topological (SPT) phases\cite{chencoho2011} of one of the emergent excitations, similar to what has been discussed for topological orders\cite{levingu}. Here we will continue this point of view  to develop a detailed understanding of the relationship between all seven phases, which is summarized in Fig.~\ref{phaserelation}. This exercise adds much new insight, and provides for new constructions of some of these phases. It also helps us obtain theories for some of the quantum phase transitions between these spin liquids.

We have already discussed how $\left(E_{fT}M_f \right)_\theta$ may be understood as a topological insulator of the $E_{fT}$ particle. So we now turn to the other phases.

Let us start from the simple $E_bM_b$ phase which can be obtained straightforwardly in microscopic models. We can obtain new phases by either putting $M$ or $E$ in {\em bosonic} topological insulator phases. $E$ transforms under $U_{eg}(1) \rtimes Z_2^T$ (meaning that the electric charge is $\T$-even) while $M$ transforms under $U_{mg} \times Z_2^T$ (meaning that the magnetic charge is $\T$-odd) where $U_{eg}(1)$ is the electric gauge transformation and $U_{mg}(1)$ is the magnetic gauge transformation.

We discuss this first for the $M$ particle. Consider {\em bosonic} topological insulators with {\em global} symmetry $U(1) \times Z_2^T$.  There are a total of $16$ such phases corresponding to classification by the group $Z_2^4$. These can be obtained from $4$ `root' phases (the 4 generators of $Z_2^4$) and taking their combinations. Two of these root phases are protected by time reversal alone while the remaining two require the full $U(1) \times Z_2^T$ symmetry. Now consider coupling these bosons to a dynamical $U(1)$ gauge field, {\em i.e} gauging the global $U(1)$ symmetry. The two root phases whose distinction requires also the $U(1)$ subgroup then potentially lead to gauge theories with distinct {\em bulk} excitations\footnote{The other phases correspond to combining the $U(1)$ liquids with SPT paramagnets protected by time reversal alone. We defer a discussion of these to Sec. \ref{addspt}}. Taken together with their combinations we get a total of four potentially distinct $U(1)$ spin liquids. The 
understanding of such bosonic SPT phases shows that these are precisely the four $U(1)$ spin liquids with a bosonic monopole ($E_bM_b$, $E_fM_b$, $E_{bT}M_b$, $E_{fT}M_b$) discussed in Sec. \ref{bosonM}.

Next consider starting with $E_bM_b$ and putting $E$ in a boson topological insulator. Such insulators with $U(1) \rtimes Z_2^T$ symmetry are classified by $Z_2^3$ with $3$ root phases. Of these only one is protected by the full $U(1) \rtimes Z_2^T$ symmetry.  Coupling the $E$, when it forms this SPT state, to a dynamical $U(1)$ gauge field  then leads to fermion statistics of the $M$ particle\cite{hmodl,metlitski}.   Thus we obtain the $E_bM_f$ spin liquid.  The fermionic statistics of the $M$ particle can be understood from a $\theta$-term in the gauge theory with $\theta=2\pi$ (Ref.~\cite{avts12}) and was called the ``statistical Witten effect''\cite{metlitski}. Likewise starting with $E_{bT}M_b$ state one can also put the Kramers bosonic charge $E$ in an SPT state and obtain the $E_{bT}M_f$ state.

Let us now understand the phases obtained by starting with  $E_bM_f$ and putting $M$ in a topological insulating phase. 
We first recall  that the monopole transforms under 
$U_{mg}(1) \times Z_2^T$ where the $U_{mg}(1)$ is the (magnetic) gauge transformation. Free fermions with {\em global } symmetry $U(1) \times Z_2^T$ can form topological band structure classified by $Z$, {\em i.e} there are distinct phases indexed by an integer $n$ which counts the number of Dirac cones at the surface. With interactions this collapses to a 
$Z_8$ classification\cite{3dfSPT2, maxvortex}.  Hence we only need to consider $n\hspace{1pt}(mod\hspace{2pt}8)$. Of these $n = 4$ is protected by $Z_2^T$ alone.  We now argue that if the global $U(1)$ is gauged, as appropriate in the $U(1)$ spin liquid, then $n = 0, 2$ (and only these) lead to distinct (at the level of bulk excitations) phases 

$n = 0$ corresponds simply to the $E_bM_f$ phase.  Interestingly $n = 2$ corresponds to the $E_{bT}M_f$ phase\cite{3dfSPT2}. 


\subsection{A puzzle}
When $n = 1$, the $U(1)$ gauge field acquires a $\theta$ term at $\theta = \pi$. This is an example of a `topological Mott insulator'  that seems distinct from the one discussed in the previous section.  In contrast to the description of the $\left(E_{fT}M_f \right)_\theta$ as a topological insulator of $E_{fT}$, here the $\theta$ term originates from the $M$ sector and leads to a `dual' Witten effect whereby the $E$ particle acquires magnetic charge $1/2$. How do we reconcile this with our claim that the list of $7$ phases is complete?

\subsection{Resolution: a dual description of the Topological Mott Insulator}

\begin{figure}
\begin{center}
\includegraphics[width=1.9in]{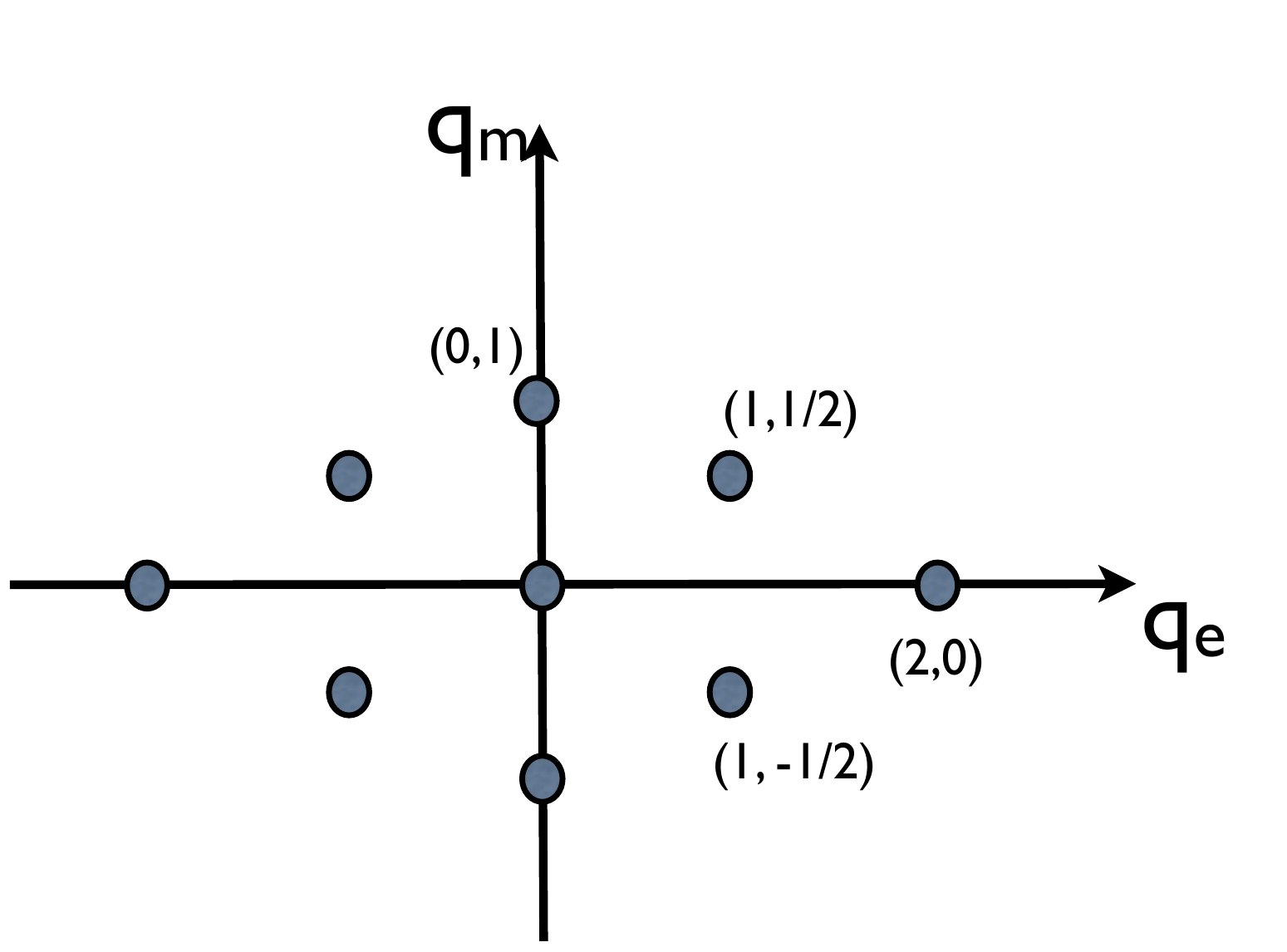}
\end{center}

\caption{ Charge-monopole lattice obtained by gauging the $n = 1$ $M_f$ topological insulator. It is identical to Fig. \ref{cmlat2} after rescaling the two axes as explained in the text.  }
\label{cmlat3}

\end{figure}

The resolution of the puzzle above is that the $n = 1$ topological insulator formed by the $M_f$ particles is actually identical to the $\left(E_{fT}M_f \right)_\theta$ phase. To see this consider the charge-monopole lattice of the $n = 1$ monopole topological insulator. This is shown in Fig. \ref{cmlat3}. Clearly it is is very similar to that of $\left(E_{fT}M_f \right)_\theta$. In Sec. \ref{lattice} we made the choice of units that the minimum pure electric charge is $1$. This leads - when there were fundamental dyons - to a minimum pure magnetic charge of $2$. We could equally well have chosen units so that the minimum pure magnetic charge is $1$. Dirac quantization (together with the existence of fundamental dyons) then would demand that the minimum pure electric charge is $2$.  This corresponds to taking the lattice in Fig. \ref{cmlat2}, shrinking the $q_m$-axis by a factor of $2$, while expanding the $q_e$-axis by a factor of $2$.  This converts the lattice in Fig. \ref{cmlat2} to that in 
Fig.~\ref{cmlat3}. 
Clearly this change in the unit choice 
does not change the physics.  In particular we correctly obtain that the pure electric charge (which has charge-$2$ in these units) is a Kramers 
fermion. 

Thus this is not a new phase but rather is included in our list. 

Let us revert back to the units where the minimum pure electric charge is $1$. We see that remarkably there are two equivalent descriptions of the $\left(E_{fT}M_f \right)_\theta$ phase which are dual to each other. We can either describe it as a topological insulator of $E_{fT}$ or as a $n = 1$ topological insulator of $M_f$.  This leads to a number of interesting consequences which will be explored in subsequent sections.



In Appendix~\ref{samegauging} we show why $M_f$ topological insulators with other values of $n$ do not lead to distinct $U(1)$ spin liquids.   {We note, and will discuss in greater detail below, that a duality similar to the one above also exists for  the spin liquid $E_{bT}M_f$.  First it can be thought of as the $n = 2$ $U(1)\times\mathcal{T}$ topological insulator of the $M_f$ particle. Equivalently it can also be  viewed  as a boson topological insulator with Kramers charge (the $E_{bT}$). This duality will have interesting consequence for the surface state, which we discuss in Sec.~\ref{tmisurf}, and for a loop wave function for this phase (Sec. ~\ref{wwloop}. }

\section{Surface states}
\label{surf}
The understanding of the connection between these $U(1)$ liquids and SPT states immediately raises the question of whether there are non-trivial surface states at the boundary between any of these spin liquids and the vacuum. In this section we discuss the necessity (or lack there of) of nontrivial surface states on general grounds. We then discuss two interesting examples: the $\left(E_{fT}M_f \right)_\theta$ and the $E_{bT}M_f$ spin liquids, which necessarily have non-trivial surface states. In Appendix~\ref{surfaceappendix} we discuss surface states of the other phases, and the interface between different phases.

\subsection{Why surface states?}
\label{whysurface}

\begin{figure}
\begin{center}
\includegraphics[width=1.9in]{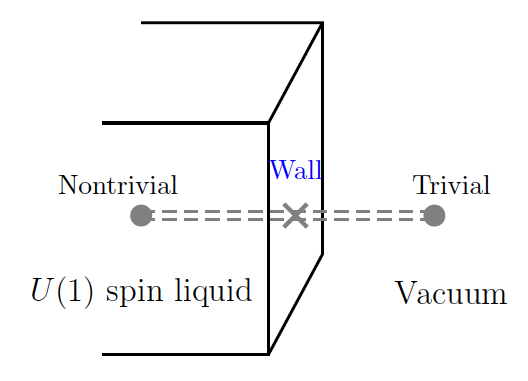}
\end{center}

\caption{The ``wall'' between a $U(1)$ spin liquid and a Higgsed vacuum. A particle ($E$ or $M$) tunnels through the wall and becomes a trivial boson, which subsequently condenses and forms the vacuum. Some nontrivial excitation must be left behind on the wall after the tunneling process.}
\label{wallfig}

\end{figure}

To see why (or why not) there should be a surface state between a $U(1)$ spin liquid and the vacuum, we should first understand what exactly is a vacuum. Since the $U(1)$ gauge field ``disappears'' in the vacuum, we should really think of the vacuum as a confined phase of the $U(1)$ gauge theory. In $(3+1)$ dimensions, a $U(1)$ gauge theory can confine in two ways: by condensing (Higgsing) either the $E$ or $M$ particle. Therefore if either the $E$ or $M$ particle is a non-Kramers boson and the vacuum is simply the condensate of that particle, the surface state will be featureless. However, if either the $E$ or $M$ particle carries nontrivial quantum number (fermion statistics or Kramers degeneracy), it cannot directly condense and form the vacuum. Instead, it should go through a ``wall'' that converts it into a trivial boson, so that the trivial boson could condense and form the vacuum. The ``wall'' then forms the surface between the $U(1)$ spin liquid and the vacuum, and obviously something nontrivial 
is needed on the wall for the conversion (Fig.~\ref{wallfig}). A similar reasoning was used in a slightly different context in Ref.~\cite{metlitski}.  We name the $E$-particle-converting wall as $E$-wall, and likewise $M$-wall for the $M$-converting wall (a similar notation was also used for $\mathbb{Z}_2$ spin liquids in $2D$, for example in Ref.~\cite{BBK}). Since the $E$-condensate and the $M$-condensate are really the same vacuum, the $E$-wall and $M$-wall can evolve into each other through phase transitions on the surface, without actually changing the vacuum.

Therefore for the two phases $\left(E_{fT}M_f \right)_\theta$ and $E_{bT}M_f$, in which all the fundamental particles are nontrivial, a nontrivial surface state is necessary, no matter how we view the vacuum. For the other five phases in Table~\ref{u1gauge1}, at least one of the $E$ and $M$ particle is a trivial boson, hence the surface is allowed to be featureless. However, they can nevertheless have nontrivial surface states if we view the vacuum differently. This will be particularly relevant if we consider the interface between two different $U(1)$ spin liquids, which we discuss in Appendix~\ref{surfaceappendix}

\subsection{Surface of $\left(E_{fT}M_f \right)_\theta$}
\label{tmisurf}

As a particularly interesting example, we now describe the surface of the $\theta = \pi$ spin liquid. Within the Abrikosov fermion slave particle construction this state   (see Ref. \cite{pesinlb} )  appears as a topological insulator of the $E_{fT}$ particle. A natural conclusion is that the surface will have an odd number of Dirac cones of the $E_{fT}$ particle which are then coupled to the bulk gapless $U(1)$ gauge field. In the language of Sec.~\ref{whysurface}, this is an $M$-wall, since a charge-$1/2$ dyon can tunnel through the wall and become a pure monopole $M$, which can subsequently condense and form the vacuum. The $M$-converting phenomenon is simply a manifestation of the parity anomaly\cite{parity} of the surface Dirac cone: a tunneling process that changes the surface flux by $2\pi$ will change the total charge\cite{kapustinqed} on the surface by $\pm1/2$.

The surface could, just like for the topological insulator, break time reversal symmetry. The Dirac fermion will then be gapped, but the surface will have a half-integer hall conductance for the gauge field. Alternately the surface could spontaneously enter a Higgs phase by condensing a bosonic Cooper pair formed out of the $E_{fT}$ particles. This corresponds to the surface superconductor, in the underlying topological insulator. Again the matter fields are gapped on the surface, but there will be vortex excitations with nontrivial fusion and braiding properties. Finally for the underlying topological insulator gapped symmetry preserving surface states with anyonic excitations are also possible\cite{fSTO1,fSTO2,fSTO3,fSTO4}. In the $U(1)$ spin liquid these anyons will, if charged, be coupled to the $U(1)$ gauge field. In all these cases, the surface states are $M$-walls.

The alternate view of this state as a monopole topological insulator leads naturally to a very different gapless symmetry preserving surface state: The $n = 1$ $M_f$ topological insulator has at the surface a single Dirac cone formed out of the $M_f$ particles. This Dirac cone is necessarily at the neutrality point as the density of $M_f$ is odd under time reversal. 
In contrast to the electric Dirac cone, tunneling a dyon through the wall gives a pure electric charge. Therefore this monopole Dirac cone is an $E$-wall, and is a very different `dual' possibility for the surface state of the topological Mott insulator. 

Given a realization of the $\left(E_{fT}M_f \right)_\theta$ phase which of these many surface phases is realized will be determined by microscopic details. As the parameters of a microscopic Hamiltonian are tuned while keeping the bulk in this phase the surface may undergo phase transitions between these various phases. In particular the $E_{fT}$ Dirac cone state may transition into the $M_f$ Dirac cone as the parameters are varied. This is related to a ``dual'' Dirac liquid state that can be realized on the surface of a topological insulator\cite{wstoappear, maxashvin}.

\subsection{Surface of $E_{bT}M_f$}
\label{btfsurface}
The $E_{bT}M_f$ $U(1)$ spin liquid is obtained naturally as a $n = 2$ $M_f$ topological insulator. This point of view immediately tells us that the surface will have a  gapless state with $2$ Dirac cones (at neutrality) of the 
$M_f$ particles. These will be coupled to the bulk $U(1)$ gauge field. In this case all $M_f$ topological insulators with $n = 2\hspace{1pt}(mod \hspace{1pt}4)$ lead to the same $E_{bT}M_f$ $U(1)$ phase in the presence of the bulk gauge field. Therefore the gapless symmetric surface state  may have $n = 2\hspace{1pt} (mod\hspace{1pt} 4)$ number of Dirac cones. This surface is an $E$-wall since it converts the Kramers $E$ particle into a non-Kramers one upon tunneling. Just as in the previous subsection alternate surface states with deconfined anyonic excitations or breaking symmetries are also possible.

The $E_{bT}M_f$ state can also have a nontrivial $M$-wall. We describe it in more detail here as an example of a surface state with gapped matter fields. It will also be useful in constructing a loop wavefunction of the spin liquid phase, which we discuss in Sec.~\ref{wwloop}.  In such a surface the electric charge is gapped and form a $\mathbb{Z}_2$ topological order, with topological quasi-particles denoted as $\{1,e,m,\epsilon\}$. The symmetries are assigned to the quasi-particles as follows: $e$ and $m$ carry electric charge $q_E=1/2$, $\T$ switches $e$ and $m$, and $\epsilon$ is charge-neutral and $\T$-singlet. Because $e$ and $m$ have a mutual $\pi$-statistics and are exchanged under $\T$, the bound state $em$ will have $\T^2=-1$. Therefore $e^2=e(m\epsilon)$ also has $\T^2=-1$, and likewise for $m^2$, which is consistent with the bulk physics since they corresponds to the charge-$1$ boson. Following the logic of Ref.~\cite{metlitski}, tunneling the $M$ through this wall will change its 
statistics from fermion to boson.

{ This surface state is precisely the surface of a boson topological insulator formed by the $E_{bT}$ particle. Such topological insulators of Kramers doublet bosons have not been discussed much in the literature (as far as we know). However their physics is easily deduced using methods developed\cite{avts12,hmodl,metlitski} for non-Kramers bosonic topological insulators.  Apart from the surface topological order described in the previous paragraph, just like their non-Kramers cousins, the $E_{bT}$ topological insulator haa a $\theta = 2\pi$ response, and associated fermionic monopoles.  }

\section{Combining $U(1)$ spin liquids with topological paramagnets}
\label{addspt}

So far we have identified two phases with the same bulk excitation spectrum as the same phase. This is reasonable if the relevant experimental probes are only detecting the bulk physics. However, additional structure needs to be considered if one is also interested in the surface states of the $U(1)$ spin liquids. In particular, one can combine a $U(1)$ spin liquid with a symmetry-protected topological (SPT) state\footnote{The meaning of `combining' is to start from two subsystems, one realizing a $U(1)$ spin liquid, the other realizing an SPT state, and couple the two systems weakly.}. This does not change the bulk spectrum in any nontrivial way, but may produce distinct surface states.

SPT states with only time-reversal symmetry are also called ``topological paramagnets''\cite{avts12}. In three dimensions, it is known that there are three nontrivial topological paramagnets. Together with the trivial state, they form a $\mathbb{Z}_2^2$ structure, which simply means combining two copies of the same state always produces a trivial state, and combining two distinct states gives the third distinct state. In the notation of Ref.~\cite{hmodl}, the three nontrivial states are labeled as 

\begin{align} \notag
&eTmT, &
& efTmfT,&
&efmf.
\end{align}

The common feature of these topological paramagnets is that they all admit gapped surface states with deconfined $\mathbb{Z}_2$ gauge theories, with topological quasi-particles labeled as $\{1,e,m,\epsilon\}$. Notice that we use small letters $e$ and $m$ to label anyons on the surface, which are not related directly to the bulk $E$ and $M$ particles. For the $eTmT$ state, both the $e$ and $m$ particles are Kramers doublets with $\T^2=-1$. For $efmf$ state, both $e$ and $m$ are fermions. The $efTmfT$ state can be viewed as the combination of the previous two states, in which both the $e$ and $m$ are Kramers fermions. The key property of these $\Z_2$ topological order is that they cannot be realized in any strictly two dimensional system while preserving $\T$. Hence they are called ``anomalous''.

The three topological paramagnets are distinct and nontrivial states when existing on their own. But do they still give distinct states when combined with a $U(1)$ spin liquid? Or equivalently, is it possible to trivialize the corresponding surface topological order in the presence of various charged matter fields? 

The stability of topological paramagnets in the presence of charged matter fields has been studied for some cases. It is known that all the topological paramagnets are stable if the charged matter field is Kramers ($\T^2=-1$), or if the $U(1)$ charge is $\T$-odd (magnetic-like). In the following we show that the $eTmT$ phase becomes trivial when the electric charge is either (a) a non-Kramers fermion or (b) a Kramers boson. These corresponds to $E_{f}M_b$, $E_{bT}M_b$ and $E_{bT}M_f$ in Table~\ref{u1gauge1}.

The argument is simple: in the presence of an electrically charged particle that is either a non-Kramers fermion or a Kramers boson, one can combine that particle with the $e$ and $\epsilon$ particle in the $eTmT$ topological order. This is essentially a relabeling of the same phase. The resulting topological order is $eCmT$, which means the $e$ particle has charge-$1$ but is non-Kramers, while the $m$ particle is Kramers but charge-neutral. This topological order turns out to be realizable even in strictly two dimensional systems. Hence it is anomaly-free. One way to realize this state is to start from the $eCT\epsilon CT$ state, which is anomaly-free since the $m$ particle is trivial, and then put the $\epsilon$ particle into a $2D$ topological insulating band. The resulting state is well known\cite{ecmt} to be the $eCmT$.

The $efmf$ state is nontrivial even in the presence of charged particles. The easiest way to see this is to notice that the $\T$-broken surface will have nontrivial thermal hall conductance, which cannot be canceled by charge matter field without introducing another hall conductance for the gauge field.

Therefore the topological paramagnets give rise to four distinct states when combined with $E_bM_b, E_bM_f, E_{fT}M_b$ and $(E_{fT}M_f)_{\theta}$, and only two distinct states when combined with the other three phases in Table~\ref{u1gauge1}. The total number of phases is thus $4\times4+3\times2=22$.

\section{Parton constructions}
\label{prtn}
We now use the insights obtained in previous sections to describe parton constructions of these $U(1)$ spin liquids. 
First let us recall that $E_bM_b$ is accessed through the gauge mean field theory, the $E_{bT}M_b$ 
through the Schwinger boson representation, and $E_{fT}M_b$ through the Abrikoson fermion representation. Further $(E_{fT}M_f)_\theta$ is accessed in the Abrikosov fermion representation by putting the fermionic spinons in a topological band insulator. 

To obtain ``natural" parton constructions for the other phases let us consider a spin-$1/2$ magnet on some lattice and use the Abrikosov fermion representation:
\begin{equation}
\label{su2pt}
\vec S_r = \frac{1}{2}f^\dagger_{r\alpha} \vec \sigma_{\alpha\beta} f_{r\beta}
\end{equation}
Here $f_{r\alpha}$ is a fermion of spin $\alpha = \uparrow, \downarrow$ at sites $r$ of the lattice. As is well known  this  representation introduces an $SU(2)$ gauge redundancy and correspondingly the physical Hilbert space of the microscopic spin system is obtained by imposing a constraint\cite{wenbook}.

It is convenient for some of our discussion to work with Majorana fermions $\eta_{ar\alpha}$ ($a = 1,2$) rather than the complex fermions $f_{r\alpha}$. We therefore define
\begin{equation}
f_{r\alpha} = \frac{1}{2}\left( \eta_{1r\alpha} - i \eta_{2r\alpha}\right) 
\end{equation}
Let $\rho^x, \rho^y, \rho^z$ be Pauli matrices acting in $\eta_1, \eta_2$ space. It is easy to check that the physical spin operators can be written as
\begin{equation}
\vec S_r = \frac{1}{8} \eta^t_r \left( \rho^y \sigma^x, \sigma^y, \rho^y \sigma^z \right) \eta_r
\end{equation}

The $SU(2)$ gauge redundancy of the fermion representation is generated by the operators 
\begin{equation}
\label{ggenerator}
\vec T = \frac{1}{8}\eta_r^t \vec I \eta_r
\end{equation}
with $\vec I = (\sigma^y \rho^x, \rho^y, \sigma^y \rho^z)$.  From these we can construct $SU(2)$ gauge transformations $O_r$ which rotate the Majorana fermions:  
\begin{equation}
\eta_r \rightarrow O_r \eta_r
\end{equation}

We first review how to obtain a $U(1)$ spin liquid through this fermionic parton construction before explaining how to implement time reversal.  We consider a mean field ansatz described by a Hamiltonian quadratic in the fermion operators: 
\begin{equation}
H_{mean} = \sum_{rr'} \eta_{r}^t h_{rr'} \eta_{r'} 
\end{equation}
with $h_{rr'}$ is a pure imaginary $4 \times 4$ matrix. Further we must have $h^t_{rr'} = - h_{r'r}$.  Under the $SU(2)$ gauge transformation, $h_{rr'}$ gets replaced by 
$O^t_r h_{rr'} O_{r'}$.

The unbroken gauge structure is determined by considering the ``Wilson loop" starting from some base point $r$
\begin{equation}
\label{wilson}
W_r[C] = \prod h_{r_i r_{i+1}}
\end{equation}
The right side is an ordered product over the $h$-matrices connecting the points $r_i, r_{i+1}$ that define the closed curve C. To get a $U(1)$ spin liquid, all the $W_r{C}$ (for different C and bases $r$) must be invariant under a $U(1)$ subgroup of the full $SU(1)$ gauge group and only this $U(1)$ subgroup.  In this case there will be a $U(1)$ subgroup of the gauge transformation 
\begin{equation}
M(\phi) = \prod_r M_r(\phi)
\end{equation}
(with each $M_r(\phi)$ describing an $SO(2)$ gauge rotation by angle $\phi$) which leaves the mean field invariant:
\begin{equation}
M_r(\phi)^t h_{rr'} M_{r'}(\phi) = h_{rr'}
\end{equation}
Let $n_r$ be the infinitesimal hermitian generators of $M_r$for each $r$ ($n_r$ will be a purely imaginary, antisymmetric  $4 \times 4$ matrix) . Upon including fluctuations  $n_r$  will correspond to the generators of  $U(1)$ gauge transformations of the spin liquid. 


Now let us consider implementation of physical global symmetries. 

In line with the rest of the paper we consider systems where time reversal is a good global symmetry. We make no assumptions about spin rotation symmetry. 
To discuss time reversal properties it is important to distinguish two distinct microscopic situations. The physical Hilbert space at each site consists of two states - these may correspond either to a Kramers doublet or to a non-Kramers doublet. Note that this distinction should not be confused with the time reversal properties (Kramers or not) of the emergent $E$ particle excitations. When the physical on-site Hilbert space corresponds to a Kramers doublet the spin operators transform under time reversal as
\begin{equation}
\label{TphysK}
 S_r^z \rightarrow -  S_r^z, ~~S_r^+ \rightarrow - S_r^-
\end{equation}

In contrast if the physical on-site Hilbert space corresponds to a non-Kramers doublet we take the spin operators to transform under time reversal as 
\begin{equation}
\label{TphysNK}
 S_r^z \rightarrow - S_r^z, ~~S_r^+ \rightarrow S_r^-
\end{equation}

For clarity we will focus henceforth on Kramers spin systems (Eq.~\eqref{TphysK}). It is straightforward to extend the discussion to non-Kraners spins. Let us now implement time reversal on the fermion operators. We may generally write 
\begin{equation}
\eta_r \rightarrow \tilde{T} \eta_r
\end{equation}
where $\tilde{T}$ is a $4 \times 4$ real matrix.  Clearly we also have the freedom to gauge transform the fermions as part of the symmetry implementation, {\em i.e} we can multiply 
$\tilde{T}$ by any gauge rotation $O_r$. 

For Kramers spins satisfying Eq.~\eqref{TphysK} we can take $f_{r} \rightarrow i\sigma^y f_r$.  This is equivalent to 
\begin{equation}
\label{TmajK}
\tilde{T} =  i\sigma^y \rho^z \eta_r
\end{equation}
 
If a mean field ansatz is time reversal invariant then we must have 
\begin{equation}
\tilde{T}^t (-h_{rr'}) \tilde{T} = O_r h_{rr'} O_{r'}^t
\end{equation}
for some gauge transformation $O_r$.  The $(-)$ sign in the left side is because time reversal is anti unitary and $h_{rr'}$ is pure imaginary.  Thus we can define a ``physical" time reversal transformation $T_r$ (for any given mean field) through
\begin{equation}
\label{tphys}
T_r = \tilde {T} O_r
\end{equation}
under which the mean field is {\em manifestly} time reversal invariant: 
\begin{equation}
T^t_r h_{rr'} T_{r'} = - h_{rr'}
\end{equation}

Now let us consider the algebra of the $U(1)$ gauge generators $n_r$ and the physical time reversal transformation.  The gauge charge $N_r$ at site $r$ is 
\begin{equation}
N_r = \eta_r^t n_r \eta_r
\end{equation}
Under time reversal we have 
\begin{equation}
\T^{-1} N_r \T^{-1} = \eta_r^t T_r^t (-n_r) T_r \eta_r
\end{equation}
The $(-)$ sign in the right side is because $n_r$ is pure imaginary. 
Generally we have 
\begin{equation}
\label{Tnr}
T^{-1}_{r} n_r T_r = \pm n_r
\end{equation}
The $(-)$ sign describes the group $U(1) \rtimes Z_2^T$, and the $(+)$ sign the group $U(1) \times Z_2^T$ (note that if $n_r \rightarrow -n_r$, then the gauge charge $N_r \rightarrow N_r$ so that the gauge charge is even under time reversal).  In the former case we should take the fermions to be the $E$ particle of the gauge theory, and in the latter we should take them to be the $M$ particle. Thus the same parton framework naturally describes both classes of phases  where $E$ is a fermion or where $M$ is a fermion.  \footnote{It is easy to show that if $n_r$ is even under time reversal at one site it must be even at all other sites that are connected to it and vice versa.}

Note that $T_r^2 = \tilde{T}O_r \tilde{T}O_r$. Even though $\tilde{T}^2 = -1$ we do not a priori know anything about $T^2$. However we know that if $n_r$ is even under $T_r$,  
then $T_r^2 = \pm 1$ (while if $n_r$ is odd we can define a modified time reversal $T_rM_r$ and $(T_rM_r)^2$ can have any value). Thus by choosing the mean field ansatz (which enables us to define $T_r$ and $M_r$) we can access phases where $E$ is either a Kramers singlet or Kramers doublet fermion. 

We now use this framework to construct examples of the $E_bM_f$, $(E_{fT}M_f)_\theta$, and $E_{bT}M_f$ phases.  For concreteness we specialize to the three dimensional cubic lattice. Consider a mean field ansatz where there is a nearest neighbor hopping $t$ and a singlet pairing $\Delta$ on the body diagonal. This corresponds to 
\begin{equation}
h_{rr'} = t_{rr'} \rho^y + \Delta_{rr'} \sigma^y \rho^x
\end{equation}
It is easy to check that the fermion spectrum is gapped. Further the non-trivial Wilson loops are proportional to $\sigma^y \rho^z$ so that this is a gapped $U(1)$ spin liquid. Correspondingly we have $M_r(\phi) = e^{i \phi \epsilon_r \sigma^y \rho^z}$ where $\epsilon_r = +1$ on the A-sublattice and $-1$ on the B-sublattice.  The physical time reversal operator can be simply taken to be $T = i\sigma_y\rho^z$. Thus the generator $\sigma^y \rho^z$ of $M_r$ is odd under $T$, and the fermions should be identified with with the $M$ particle. Further it is also readily checked that the band structure is not topological. We thus have a realization of the $E_bM_f$ phase. 

Next let us modify the $t$ and $\Delta$ to get a topological band structure.  Precisely such a modification was discussed in Ref. \cite{hosur}, and requires changing the sign of the $t$ and $\Delta$ on some of the bonds. This yields a $n = 2$ topological insulator with two surface Dirac cones.  The $U(1)$ gauge structure and time reversal properties are not affected by this modification. We thus end up with the $E_{bT}M_f$ phase. 

Finally, to construct the $\left(E_{fT}M_f\right)_\theta$ phase we use a different implementation of time reversal. We take $\eta_r \rightarrow \epsilon_r \eta_r$ corresponding to $T_r = \epsilon_r$.  We take a band structure in which the two $\eta_{\uparrow}$ fermions (which make up the complex fermion $f_\uparrow$) have different dispersion than the $\downarrow$ fermions. Specifically we choose the $\uparrow$ band structure described in Ref. \cite{hosur} for the $n = 1$ topological insulator while for the $\downarrow$ we choose a trivial dispersion\footnote{This band structure is invariant under the chosen time reversal operation}.  It is easy to check that this mean field ansatz describes a $U(1)$ spin liquid, and further that the $n_r$ are odd under time reversal. Thus the fermions must be identified with $M$. Further as we have a net $n = 1$ topological band structure we get the 
$\left(E_{fT}M_f\right)_\theta$ phase. 

We emphasize that this construction is totally different from the ``standard"  one where the fermions are treated as Kramers doublet $E$ particles with topological band structure. Nevertheless we get the same phase.

\subsection{Kramers spin on non-bipartite lattice}
\label{nogo}

\subsubsection{$M_f$-no-go}
\label{mfnogo}

The previous examples of $M_f$-type parton construction, in which the fermions are monopole-like with their $U(1)$ gauge charge odd under $\T$, were constructed on a bipartite lattice (cubic lattice). We now show that for Kramers spins (Eq.~\eqref{TphysK}) such $M_f$-type construction is impossible on a non-bipartite lattice.  First we note that in Eq.~\eqref{Tnr}
we must choose the $+$ sign in this case.  Second we notice that  $\tilde{T} = i\sigma^y \rho^z$ is itself a gauge rotation. Thus the physical time reversal matrix $T_r = \tilde {T} O_r$ is also just an $SU(2)$ gauge rotation. 

Thus we may write $T_r=e^{i\theta_r \tau_r}$, where $\tau_r$ is a hermitian generator. In general $\tau_r$ is a combination of the three generators in Eq.~\eqref{ggenerator}, satisfying $\tau_r^2=1$ and $\tau_r^*=-\tau_r$.

With the $+$ sign in Eq.~\eqref{Tnr}, we have $[T_r, n_r] = 0$.  This is possible only if $n_r = \pm  \tau_r$.  For any  Wilson line with base $r$ (see Eq.~\eqref{wilson}) satisfies 
\begin{equation}
M_r^t(\phi) W_r[C] M_r (\phi)= W_r[C]
\end{equation}
As $T_r$ corresponds to a special value of $\phi$ we also have
\begin{equation}
T_r^t W_r[C] T_r = W_r[C]
\end{equation}
However using $T_r^t h_{rr'} T_{r'} = - h_{rr'}$ we can also conclude that for a loop of length $L$ 
\begin{equation}
T_r^t W_r[C] T_r =  (-1)^L W_r[C]
\end{equation}
We thus conclude that $L$ must be even which is possible only if the lattice is bipartite. A related argument using the trace of the Wilson loop to diagnose time-reversal breaking was presented in Ref.~\cite{lietal}.

This shows that for Kramers spins on non-bipartite lattices (such as the pyrochlore), fermionic monopole does not arise within the particular (although most common) type of parton construction from Eq.~\eqref{su2pt}. We should emphasize that this does not rule out the possibility of having such phases in this situation, since one can imagine having more complicated types of parton construction. However this does suggest that states with fermionic monopoles are less natural in  these systems.

\subsubsection{$E_f$-no-go}

Following the same logic, we now show that $E_f$-type parton construction, in which the fermions are electron-like but non-Kramers, is also impossible for Kramers spins on non-bipartite lattices. To have $T_r^2=1$ on the $\eta$ fermions, the only possibility is $T_r=\pm1$.  Clearly $W_r[C]$ must be real to preserve time-reversal. But as we discussed above in Sec.~\ref{mfnogo}, on a non-bipartite lattice there must exist imaginary Wilson loops. Therefore such a construction is impossible.

Again we emphasize that this does not rule out the $E_fM_b$ state, but does make it less natural in non-bipartite Kramers spin systems.

\section{Loop Wavefunctions}
\label{loops}
It is interesting to understand the differences between these different states in terms of their ground state wave functions.  To that end it is useful to think of the $U(1)$ spin liquid  in terms of 
fluctuating loop configurations. As matter fields (the $E$ and $M$ particle) are gapped, the low energy physics is described by Maxwell electrodynamics. The emergent electric and magnetic fields are divergence-free and hence the corresponding field lines form closed loops. To describe the wave function we can choose either the electric picture or the magnetic picture (these are different bases for the low energy Hilbert space). 

In the specific context of quantum spin ice the magnetic flux loops are very easy to picture. Indeed the spin ice manifold is parametrized in terms of closed loop configurations formed by the directions of the microscopic spins  on the pyrochlore lattice. Quantum effects introduce fluctuations of these magnetic loops and, in the spin liquid, lead to tensionless fluctuating loops in the ground state. 
The simplest possibility is that the wavefunction of the fluctuating magnetic loops is positive definite:
\bea
\label{ebmbloop}
|\Psi\ra&=&\sum_{\mathcal{C}}\Psi_0(\mathcal{C})|\mathcal{C}\ra, \nn
\Psi_0(\C)&\sim& e^{-\int d^3xd^3x'\alpha\frac{\vec{B}(\vec{x})\cdot\vec{B}(\vec{x}')}{|\vec{x}-\vec{x}'|^2} },
\eea
where $\alpha$ is a positive consitant and $\vec{B}$ is the magnetic field corresponding to the magnetic loop configuration $\C$. The positive weight $\Psi_0$ is needed to satisfy Maxwell's equation. Such a ``featureless'' wavefunction would describe the $E_bM_b$ phase, as studied in many previous works. 
The $M$ particle is the open end of the magnetic loops, and the $E$ particle is a point defect with an additional phase factor in the wavefunction:
\be
|\Psi(E)\ra=\sum_{\mathcal{C}}e^{i\sum_I\Omega_I/2} \Psi_0(\mathcal{C})|\mathcal{C}\ra,
\ee
where $I$ lables each loop in the configuration $\C$, and $\Omega_I$ is the solid angle spanned by the loop $I$ with respect to the $E$ particle.

To describe the other six phases in Table~\ref{u1gauge1}, more subtle structures are needed in the loop wavefunction. 
For the $E_bM_f$ phase, the monopoles--end points of the magnetic loops--need to become fermions. This can be done by thickening the magnetic loops into ``ribbons'', and assigning a phase $(-1)$ to the wavefunction whenever a ribbon self-links. More precisely, the wavefunction can be written as
\be
|\Psi\ra=\sum_{\mathcal{C}}(-1)^{L^S_{\C}}\Psi_0(\mathcal{C})|\mathcal{C}\ra,
\ee
where $L^S_{\C}$ is the self-linking number, defined to be the linking number of the two boundary loops of each magnetic ribbon.  An argument in Ref. \cite{xuts13} shows that due to this extra phase  the open end points of such loops have fermi statistics.

To understand some of the other phases described in Table~\ref{u1gauge1} in terms of fluctuating loops it is more convenient to use instead the `electric' picture: the ground state is then a superposition of oriented loops (which represent the electric field lines) with weights derived from the Maxwell action $\Psi_0\sim  e^{-\int d^3xd^3x'\alpha'\frac{\vec{E}(\vec{x})\cdot\vec{E}(\vec{x}')}{|\vec{x}-\vec{x}'|^2} }$. In the $E_bM_b$ phase these electric loops are featureless, and the superposition has positive definite weights for all loop configurations just as in Eq.~\eqref{ebmbloop}. 

In the $E_{bT}M_b$ phase, we can think of the electric field lines as `stuffed' with $1d$ Haldane/AKLT chains. One way to do it is to consider an additional spin-like order parameter $\hat{n}$ in the disordered paramagnetic phases, and assign a Wess-Zumino phase factor in the wavefunction:
\be
|\Psi\ra=\sum_{\mathcal{C}}e^{iW[\hat{n}(\C)]}\Psi_0(\mathcal{C})|\mathcal{C};\hat{n}\ra.
\ee
The easiest way to picture the Wess-Zumino term is to view $W[\hat{n}(\C)]/2\pi$ as the total skyrmion number of $\hat{n}$ on the membranes whose boundaries are the electric loops $\C$. 
For a closed loop this internal structure has no serious effect. However if we produce an electric charge we expose an open end of the electric field line. The Kramers doublet known to be present at the open end  of the Haldane chain then leads to the Kramers degeneracy of the electric charge. 
Notice that this possibility is meaningful because the electric loop configurations are time-reversal invariant. In contrast, the magnetic loops cannot be stuffed with Haldane chains, in line with the discussions in the rest of the paper.  

If instead the $E$ particle is a fermion (as in $E_fM_b$ or $E_{fT}M_b$) then the electric field is best thought of as a thin ribbon ({\em i.e} a line with some small but non-zero thickness). Again we assign a phase $(-1)$ to an electric field loop which has an odd self-linking number, which converts the $E$ particle into a fermion.  For the $E_{fT}M_b$ phase, in addition, these electric loops must be stuffed with Haldane chains. 

In all the examples above, at least one of the $E$ and $M$ particles is trivial (bosonic and non-Kramers). This makes it simpler to describe the wavefunction by considering the loops with nontrivial open ends. For example, if the $E$ ($M$) particle is trivial, we can write the wavefunction as fluctuating $M$ ($E$) loops and demand that they have the right structure to produce nontrivial end-points, which are the $M$ ($E$) particles.

The remaining two phases in Table.~\ref{u1gauge1} ($E_{bT}M_f$ and $(E_{fT}M_f)_{\theta}$), however, cannot be easily understood using the above line of thinking because both $E$ and $M$ particles are nontrivial. The loop wavefunctions for these two phases should capture not only the quantum numbers of the end points, but also that of the dual particles. Similar issue arises if we want to understand the previous phases in the dual loop picture. For example, can we have a magnetic loop wavefunction for $E_{fT}M_b$ state?

This issue is actually closely related to the surface states of the phases: if a $U(1)$ spin liquid phase necessarily has a nontrivial surface state, one should be able to infer it from the bulk wavefunction. 
The two phases $E_{bT}M_f$ and $(E_{fT}M_f)_{\theta}$ both have nontrivial surface states as long as time-reversal is kept. For the other phases like $E_{fT}M_b$, the surface has to be nontrivial as long as time-reversal is kept and the $M$ particle is not condensed on the boundary. Since the $M$ particle is naturally not condensed in the fluctuating magnetic loop picture, the wavefunction of magnetic loops should contain the information of the nontrivial boundary theory. Similar logic also applies to the other phases except $E_bM_b$.

Notice that when the wavefunction is written solely in terms of closed loops, the matter fields are naturally gapped, even on the boundary. Therefore to have loop wavefunctions for the above nontrivial cases, we need the loop structures to be able to produce boundary theories with gapped matter fields. Since the gapped matter fields on a nontrivial boundary is necessarily fractionalized, this suggests the bulk wavefunction should be described in terms of fractional loops, instead of the ``physical'' loops such as the $2\pi$ magnetic loop. 

A similar problem was tackled in the context of symmetry-protected topological (SPT) phases using what is known as Walker-Wang construction\cite{walkerwang,ashvinbcoh,fidkowski3d,chenanomaloussymm,fSTO3}. The essential idea is that when the boundary is a gapped topological order, one can have a loop wavefunction for the bulk, for which the weights are knot invariants of the loop configurations generated by the boundary topological field theory. With some modification, this 
idea can be used to generate loop wavefunctions of 
$U(1)$ spin liquids (other than the simple $E_bM_b$) in Table.~\ref{u1gauge1}. In Sec.~\ref{wwloop} we discuss a relatively simple yet interesting loop wavefunction of the $E_{bT}M_f$ phase as an illustrating example. With different time-reversal implementation the same wavefunction can also describe two other phases, namely the $E_bM_f$ in the electric loop picture, and $E_fM_b$ in the magnetic loop picture, which we describe in Sec.~\ref{wwloop2}. In Appendix~\ref{wwloopappendix} we discuss a slightly different wavefunction that can describe the $E_{bT}M_b$ and $E_{fT}M_b$ phases. The topological Mott insulator $(E_{fT}M_f)_{\theta}$ can also be described through gauging its (non-Abelian) Walker-Wang wavefunction\cite{fSTO3}, but it will be quite complicated and not very illuminating, so we will omit the discussion in this paper.

\subsection{A loop wavefunction for the $E_{bT}M_f$ phase}
\label{wwloop}

\begin{figure}
\begin{center}
\begin{subfigure}{.4\textwidth}
\includegraphics[width=2in]{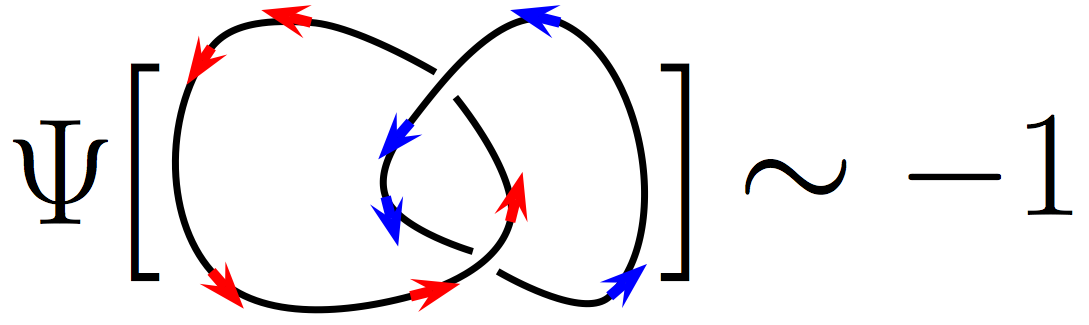}
\caption{}
\label{wf1}
\end{subfigure}
\end{center}

\begin{subfigure}{.2\textwidth}
\includegraphics[width=1.6in]{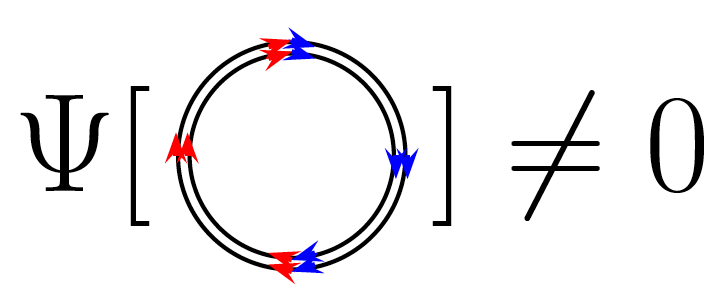}
\caption{}
\label{wf2}
\end{subfigure}
\begin{subfigure}{.2\textwidth}
\includegraphics[width=.7in]{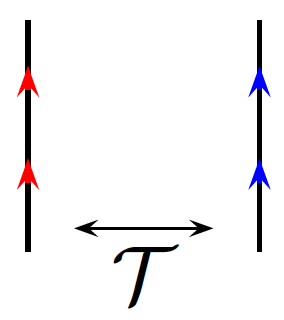}
\caption{}
\label{wf3}
\end{subfigure}
\caption{Electric-like loop wavefunction of the $E_{bT}M_f$ $U(1)$ spin liquid:  (a) The amplitude changes sign whenever a blue loop links with a red one. (b) A doubled blue line can be converted to a doubled red line and form a doubled two-segment loop. ($c$) The red and blue loops are switched under time-reversal $\T$. }
\label{linkwavefunction}

\end{figure}

For the $E_{bT}M_f$ phase, it turns out to be slightly easier to describe the wavefunction in terms of electric loops. The wavefunction is written in terms of two species of oriented loops, labeled as ``red'' ($r$) and ``blue'' ($b$):
\be
|\Psi\ra=\sum_{\mathcal{C}_r,\C_b}(-1)^{L_{\C_r,\C_b}}\Psi_0(\mathcal{C})|\mathcal{C}_r,\C_b\ra,
\ee
where $L_{\C_r,\C_b}$ is the mutual linking number between red and blue loops. Two additional features are present in the wavefunction. First, a doubled blue line can be converted to a doubled red line and form a doubled two-segment loop, even though single lines cannot be converted to each other. Second, the red and blue loops get switched under time-reversal action $\T$, which is allowed since these are electric-like loops. These features are illustrated schematically in Fig.~\ref{linkwavefunction}. 

To see that the wavefunction described in Fig.~\ref{linkwavefunction} indeed describes the $E_{bT}M_f$ phase, we need to examine the excitation spectrum. First of all, it is useful to examine the bound state of one blue loop and one red loop, with opposite directions. We call this the $\epsilon$ loop. Notice that due to the condition in Fig.~\ref{wf2}, the $\epsilon$ loop is undirected. We also note that the $\epsilon$ loop has a linking sign with both blue and red loops. This makes the end points of an individual open blue/red line confined, in the sense that they cost an energy proportional to the length of the open line. The reason is that with an open blue/red line, a small $\epsilon$ loop surrounding the interior part of the line locally behaves as though it is ``linked'' with the line. However, it cannot have a linking sign in the wavefunction. This is because one can continuously move the small $\epsilon$ loop away from the open line till it looks ``unlinked''. Therefore the local Hamiltonian near the 
interior of the open 
line cannot be minimized. Thus the energy penalty will be proportional to the length of the line. The same physics also appears in Walker-Wang models. (See also Ref.~\cite{WNS} for a simple and concrete model illustrating this.)

On the other hand, a doubled red loop (or equivalently, a doubled blue loop) can be opened with finite energy cost, since it does not have linking sign with anything. Therefore we interpret the end points of doubled red loops as the deconfined $E$ particles. The single red/blue loops are ``half'' electric loops. 

The open end of a bare $\epsilon$ line is confined because it has linking signs with the red/blue loops. However, a monopole-like defect can be bound at the end of an $\epsilon$ line to avoid the sign ambiguity and make it deconfined. More precisely, we can have a phase factor in the wavefunction of an open $\epsilon$ line $e^{i\sum_{I}(\Omega^a_I-\Omega^b_I)/4}$, where $I$ denotes all the red and blue loops, and $\Omega^{a,b}_I$ is the solid angle spanned by a red/blue loop with respect to the end points $a,b$. This phase factor serves as a smooth interpolator between the linking phase away from the line ($+1$) and the linking phase near the interior of the line ($-1$). Since the single red/blue loops are interpreted as half electric loops, such a phase factor corresponds precisely to a magnetic monopole with unit magnetic charge. 

Therefore the magnetic monopole is bounded to the end point of an $\epsilon$ line. But notice the $\epsilon$ loop is really a ribbon, with a red and a blue loop being the edges of the ribbon. Therefore it has a self-linking sign from the red/blue mutual linking sign, which makes the end point a fermion. We have thus obtained fermion statistics of the magnetic monopole!

We now discuss time-reversal action on the $E$ particle, which is the end point of a doubled blue line. For this purpose it is convenient to view the doubled blue line as the combination of a blue, a red and an $\epsilon$ line. Under time-reversal $\T$ the $\epsilon$ line is invariant, but the blue and red lines are exchanged (see Fig.~\ref{wf3}). Since the time-reversed wavefunction away from the charged particle is locally indistinguishable from the original wavefunction, the $\T$ action can be effectively ``localized'' around the $E$ particle, by exchanging the two end points of the red and blue lines. Therefore performing $\T$ twice amounts to twisting the blud-red ribbon, which gives a $(-1)$ phase in the wavefunction. This implies that the $E$ particle has $\T^2=-1$ and is a Kramers doublet.

Similar to the Walker-Wang construction, the above discussion is closely related to a possible surface state of the $E_{bT}M_f$ phase that is gapped but breaks no symmetry. This is exactly the $M$-wall state discussed in Sec.~\ref{btfsurface}.

\subsection{Alternative loop wavefunctions for $E_bM_f$ and $E_fM_b$}
\label{wwloop2}

The same wavefunction described in Fig.~\ref{linkwavefunction} can also describe two other phases, with different ways of implementing time-reversal symmetry. 

\begin{figure}

\begin{subfigure}{.2\textwidth}
\includegraphics[width=.7in]{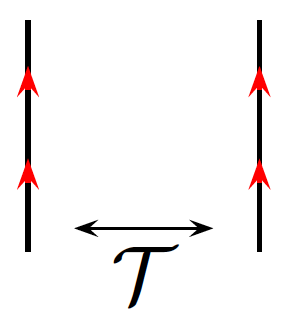}
\caption{}
\label{wf4}
\end{subfigure}
\begin{subfigure}{.2\textwidth}
\includegraphics[width=.7in]{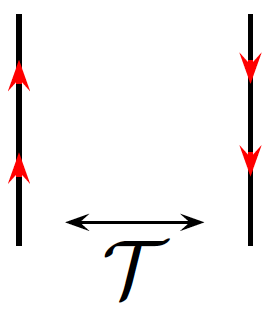}
\caption{}
\label{wf5}
\end{subfigure}
\caption{The same loop wavefunction as in Fig.~\ref{linkwavefunction}, but with different time-reversal actions. (a) Time reversal keeps both the color and the direction of each loop, which describes $E_bM_f$ in the electric picture. (b) Time reversal keeps the color, but invert the direction of each loop, which describes $E_fM_b$ in the magnetic picture.}
\label{linktr}

\end{figure}

If time-reversal $\T$ keeps both the color and the direction of each loop (Fig.~\ref{wf4}), then the loops are electric-like. The argument in Sec.~\ref{wwloop} for the fermionic monopole still applies in this case. But the electric charge -- the end point of a doubled red (or blue) line -- now transforms trivially under time-reversal. Hence we obtain the $E_bM_f$ state in the electric loop picture.

Now if time-reversal keeps only the color, but invert the direction of each loop (Fig.~\ref{wf5}), the loops are magnetic-like. The argument in Sec.~\ref{wwloop} for the fermion statistics of the $\epsilon$ particle still applies, but now this particle should be interpreted as the electric particle $E$. We therefore obtain the $E_fM_b$ state in the magnetic loop picture.

\section{Quantum phase transition between two distinct $U(1)$ spin liquids}
\label{transitions}

In this section, we consider quantum phase transitions from one $U(1)$ spin liquid to another in Table~\ref{u1gauge1}. In general, one may expect most of the transitions to be first order or to go through an intermediate phase, due to lack of any obvious order parameter. Thus any continuous transition between two such phases would be quite exotic. The understanding of these spin liquids in terms of gauged SPT states shed some new light on this subject. If we can understand possible continuous transitions between different SPT states, we can then couple these critical theories to $U(1)$ gauge fields and understand continuous transitions between $U(1)$ spin liquids. The common feature of such phase transitions is that one particle (say $E$) is unchanged across the transition, while the dual particle (say $M$) drastically changes its properties such as statistics, $\T^2$ value, and dyon charge.

Continuous transitions between SPT phases in free fermions are well understood\cite{titransition}. Interestingly, by gauging such transitions we can already obtain many novel transitions between various $U(1)$ spin liquids. All such critical theories are described by massless Dirac fermions in $(3+1)$ dimensions coupled with a $U(1)$ gauge field, but the effect on the dual particles are very different. We are thus able to provide remarkably simple descriptions of some highly non-trivial continuous phase transitions between distinct $U(1)$ spin liquids. For instance we provide a theory for a continuous phase transition between the topological Mott insulator and the $E_bM_f$ phase. In the conventional picture of the topological Mott insulator as a spinon topological insulator, such a transition seems to require a change of statistics of the electric charge! Such a `statistics' changing quantum phase transition is however very simply understood within the dual picture of the topological Mott insulator (as a 
monopole topological insulator) developed in this paper. 

\subsection{``Statistics-changing" quantum criticality: Phase transitions of the topological Mott insulator}

\subsubsection{Warm-up: To $E_{fT}M_b$}
It is useful to first understand the phase transition from the topological Mott insulator -- the $(E_{fT}M_f)_{\theta}$ phase -- to the $E_{fT}M_b$ phase.  This will set the stage for the more surprising (from the conventional viewpoint) phase transitions studied below. 
Since $(E_{fT}M_f)_{\theta}$   can be viewed as a gauged version of a topological insulator formed by the $E_{fT}$ particles, we can access its transition into the $E_{fT}M_b$ phase by gauging the topological-to-trivial insulator transition. The critical theory is simply a massless $QED$ with one flavor:
\be
\label{qed4}
\mathcal{L}[\psi,\bar{\psi},a_{\mu}]=\bar{\psi}(i\slashed{\partial}+\slashed{a})\psi+im\bar{\psi}\psi+\mathcal{L}_{Maxwell}[a_{\mu}],
\ee 
where $\gamma_0=\tau_1$, $\gamma_i=\sigma_i\tau_2$, $\bar{\psi}=i\psi^{\dagger}\gamma_0$, and time reversal acts as $\T\psi\T^{-1}=i\sigma_2\psi$ (here $\sigma_i$ and $\tau_i$ are Pauli matrices). One can easily check that the $i\bar{\psi}\psi$ term is the only $\T$-symmetric mass term. Under a proper UV background, $m>0$ gives a trivial insulator, which after gauging becomes the $E_{fT}M_b$ phase, and $m<0$ gives the topological insulator, which after gauging becomes the $(E_{fT}M_f)_{\theta}$ phase. $m=0$ is thus the critical point.

However, to make the transition really continuous, we should also forbid other $\T$-invariant terms that close the fermion gap before the mass $m$ becomes zero. There are two such terms: $\mu\psi^{\dagger}\psi$ and $\mu'\psi^{\dagger}\tau_3\psi$. The former is simply the chemical potential term, and the latter can be viewed as a chemical potential alternating in sign for the two Weyl fermions. The chemical potential is forbidden since the total gauge charge should be zero (assuming the fermion gap does not close elsewhere). But to forbid the $\mu'\psi^{\dagger}\tau_3\psi$ term, more symmetry is required in the theory. The simplest possibility is to demand inversion symmetry $\I\psi(\vec{x})\I^{-1}=\tau_1\psi(-\vec{x})$, as was shown in Ref.~\cite{titransition}

In general, various anisotropy terms are also allowed. These include the spatial anisotropy, the velocity difference between different bands, and the difference between the fermion velocity and the speed of the emergent photon. We will not go into those details.

\subsubsection{Topological Mott insulator to $E_bM_f$}
\label{thetatoebmf}

As discussed in previous sections, the $(E_{fT}M_f)_{\theta}$ phase can also be viewed as a ``dual'' topological insulator of the fermionic monopoles. This makes it possible to access its transition to the $E_bM_f$ 
phase, in which the fermionic monopoles form a trivial insulator. The critical theory has the same Lagrangian as Eq.~\eqref{qed4}, but with a different implementation of time-reversal symmetry: $\T\psi\T^{-1}=i\tau_2\psi^{\dagger}$.

Again we need to forbid $\T$-invariant terms that can make the fermions gapless. There is only one such term, namely the alternating chemical potential $\mu'\psi^{\dagger}\tau_3\psi$. We can again forbid it by having inversion symmetry  $\I\psi(\vec{x})\I^{-1}=\tau_1\psi(-\vec{x})$.

\subsubsection{Topological Mott insulator to $E_{bT}M_f$}
 \label{thetatoebtmf}

The previous transition into $E_bM_f$ phase can be viewed as a transition of the fermion monopoles from an $n=1$ band to an $n=0$ band. Now if we consider a transition from an $n=1$ band to an $n=2$ band, this gives a transition from the topological Mott insulator to the $E_{bT}M_f$ phase. The gapless part of the critical theory is the same as the previous transition described in Sec.~\ref{thetatoebmf}. The only difference is that here we have an extra gapped Dirac fermion with negative mass, which gives the $n=1$ band in the UV background. This UV background is important in determining the nature of the phases away from the critical point. But it will not affect the physics right at the critical point such as the scaling relations. One can see this by integrating them out, which gives a $\theta$-term in the $U(1)$ gauge field. But since the dual particle (charge) is gapped at the critical point, the $\theta$-term is a purely surface term and will not affect the bulk dynamics.

\subsection{Kramers-changing quantum criticality: Phase transition between $E_{bT}M_f$ and $E_bM_f$ spin liquids}
The $E_{bT}M_f$ and $E_bM_f$ spin liquids differ by whether or not the $E$ particle is a Kramers doublet. We now discuss the criticality associated with a continuous transition where this Kramers-ness changes. 
The $E_{bT}M_f$ phase can be viewed as fermionic monopoles in an $n=2$ band. We have discussed the transition from the $E_{bT}M_f$ phase to the topological Mott insulator in Sec.~\ref{thetatoebtmf}. We now discuss the transition to the $E_{b}M_f$ phase, which is a trivial insulator ($n=0$) of the fermion monopoles.

The critical theory has two mass Dirac fermions coupled with a $U(1)$ gauge field:

\be
\label{qed42}
\mathcal{L}[\psi_s,\bar{\psi}_s,a_{\mu}]=\sum_{s=1,2}\bar{\psi}_s(i\slashed{\partial}+\slashed{a})\psi_s+im\bar{\psi}_s\psi_s+\mathcal{L}_{Maxwell}[a_{\mu}],
\ee 
where $\gamma_0=\tau_1$, $\gamma_i=\sigma_i\tau_2$, $\bar{\psi}_s=i\psi^{\dagger}_s\gamma_0$, and time reversal acts as $\T\psi_s\T^{-1}=i\tau_2\psi^{\dagger}_s$ (here $\sigma_i$ and $\tau_i$ are Pauli matrices). Again we need inversion symmetry $\I\psi_s(\vec{x})\I^{-1}=\tau_1\psi_s(-\vec{x})$ to forbid the alternating chemical potential term $\mu'\psi_s^{\dagger}\tau_3\psi_s$. Moreover, we need $im\bar{\psi}_s\psi_s$ to be the unique mass term. This requires some kind of rotation symmetry between the two flavors $s=1,2$. Microscopically this could be achieved with certain lattice symmetries. We will not go into such details.

\section{Role of time reversal} 
We should emphasize that the seven phases are distinct if and only if  time-reversal symmetry is kept. In the absence of time-reversal, $\T^2$ of course has no meaning, and the $\theta$-angle can be continuously tuned to any value. Even different statistics of particles do not distinguish phases: if the magnetic particle is a fermion, the electric particle is necessarily a boson, and one can change the $\theta$-angle by $2\pi$ to shift the $m$ particle to a boson\cite{metlitski}. Similar argument applies if the electric particle is a fermion.

Thus in the absence of any symmetries we have exactly one $U(1)$ liquid phase.

 

 \section{Relationship with other works}
We have focused on time reversal symmetric $U(1)$ quantum spin liquids. We now place our results in the broader context of research on symmetry implementation in other long range entangled phases. 
 The best understood long range entangled phases have a gap to all bulk excitations, and are characterized by the concept of topological order. In the last few years the realization of symmetry in such topologically ordered phases has gotten a great deal of attention. Such states have been dubbed `Symmetry Enriched Topological' (SET) phases.   As is well known, in an SET phase  the topologically non-trivial quasiparticles may carry fractional quantum numbers.  This  means that the action of symmetry on these quasiparticles is non-trivial (technically the symmetry is realized projectively rather than linearly). The projective realization is allowed since these quasiparticles are non-local objects. Symmetry operations may also have more dramatic effects: they may even interchange two different topological sectors. 
 
 In $d = 2$ space dimensions there is significant progress in classifying and understanding these SET phases.  For some representative papers see Refs.~\cite{Levin,Santos,essinh,ran12,janet13,geraetds,luav,cho,kapustinthorngren,barkeshli,teo} and \cite{avts12,hmodl,chenanomaloussymm}. 
 For the three dimensional systems of interest in this paper, some progress in understanding gapped time reversal symmetric $Z_2$ quantum spin liquids has been reported in Ref. \cite{xu3dz2}.  The $U(1)$ spin liquids discussed in this paper are gapless but nevertheless we have shown how we can classify and understand the realization of time reversal symmetry.

 We have already discussed previous model constructions of some of these phases. Microscopic models for $E_bM_b$ are common\cite{bosfrc3d,wen03,hfb04,3ddmr,lesikts05,kdybk,shannon}.  What about the other phases? 
 Ref. \cite{levinu1f} constructed a rotor model in which one of the emergent particles is a fermion. This can be viewed as either a construction of $E_fM_b$ or of $E_bM_f$ depending on how time reversal is implemented on the microscopic rotor degrees of freedom.  If we take it to be a model for $E_bM_f$ it should be possible to modify the $M_f$ hopping to give it topological band structure. This will enable writing down microscopic models for $(E_{fT}M_f)_{\theta}$ and $E_{bT}M_f$. We will however not pursue this here. 
 
 \section{Discussion: models, materials, and experiments}
\label{mat}
We now consider the lessons learnt from our results for current and future possible experimental realizations of $U(1)$ spin liquids. We discuss two separate issues. First, for a given system, if a $U(1)$ spin liquid arises, which of the seven families of phases in Table~\ref{u1gauge1} is realized? This is particularly relevant to the quantum spin ice materials such as the pyrochlore $Yb_2Ti_2O_7$. Existing theoretical work\cite{gMFT} assumed that the simplest phase in the $E_bM_b$ family is the prime candidate in such systems. However this is justified only deep in the spin ice limit\cite{hfb04}, and materials such as $Yb_2Ti_2O_7$ are quite far away from this limit (see review in Appendix~\ref{qspicemdl}). So it is important to ask which of the seven phases discussed in this paper is more likely to arise in such systems. We address this issue in Sec.~\ref{pyro}.

The next important issue is to identify distinguishing features of these different phases that can be probed in experiments. We partially address this issue in Sec.~\ref{experimental}.

\subsection{Pyrochlore spin ice}
\label{pyro}

Here we will focus on Kramers spin systems since they tend to be more robust against disorder. It is of course very hard to decide energetically which phase is more favorable, due to the complexity of the underlying Hamiltonian. But at least we can ask the following question: which of the seven phases in this paper have a natural mean-field description on the pyrochlore lattice? Clearly the gauge mean field theory (gMFT) proposed in Ref.~\cite{gMFT} is a natural mean field theory for the phase $E_bM_b$. So what about the other six phases?

As discussed in Sec.~\ref{nogo}, for Kramers spins on a non-bipartite lattice such as the pyrochlore, it is quite unnatural - at the level of parton mean field theory - to have   fermionic monopoles or non-Kramers fermionic electric charges.  This is already enough to render unlikely the  $E_fM_b$, $E_bM_f$ and $E_{bT}M_f$ phases. 

The phase $E_{bT}M_b$ is also unnatural at the mean field level on a non-bipartite lattice. To construct it using mean field, we need to use the Schwinger boson decomposition $S_{\mu}=\frac{1}{2}b^{\dagger}_{\alpha}\sigma_{\mu}^{\alpha\beta}b_{\beta}$. The mean field theory should have one boson per site on average, and the bosons need to be gapped. On a non-bipartite lattice this is impossible at the mean field (quadratic) level, and the bosons will always tend to either condense or pair condense, which breaks the $U(1)$ gauge symmetry.

We are thus left with only two phases: the $E_{fT}M_b$ and $(E_{fT}M_f)_{\theta}$. Both can be described at the mean field level through the Abrikosov fermion decomposition $S_{\mu}=\frac{1}{2}f^{\dagger}_{\alpha}\sigma_{\mu}^{\alpha\beta}f_{\beta}$, where time-reversal acts as $f\to i\sigma_yf$, and the $U(1)$ gauge symmetry is the phase rotation on the fermions: $f_{\alpha}\to e^{i\theta}f_{\alpha}$. 

We can try to write down mean field band structures of these $f_{\alpha}$ fermions on the pyrochlore lattice. We restrict to mean field Hamiltonians that has only nearest-neighbor terms, which is reasonable since the spin exchange is very short-ranged in quantum spin ice materials. We further restrict to mean field Hamiltonians that are manifestly invariant under the full lattice symmetry. This is less justified, since the $f_\alpha$  generally are allowed to transform projectively under the symmetry group. Nevertheless we make this assumption in order to make progress, while keeping this caveat in mind. 

With these restrictions, the mean field Hamiltonian has only two parameters: the trivial hopping term $t_1$ and the spin-orbit coupled hopping term $t_2$ and takes the form
\begin{equation}
\label{mfhopping}
H_{MF} = -\sum_{\langle rr' \rangle} f^\dagger_r \left(t_1 + i t_2 \vec d_{rr'} \cdot \vec \sigma \right) f_{r'}
\end{equation}
Here $\vec d_{rr'}$ is a unit vector parallel to the opposite bond of the tetrahedron containing ${rr'}$ (see Ref.~\cite{william} for details). 
The resulting band structure has been studied elsewhere, for example in Ref.~\cite{william}. The amusing fact about this Hamiltonian is that as long as it is gapped, the fermions will always form a topological insulator! We therefore reach the conclusion that, beside the $E_bM_b$ state described by gMFT, the topological Mott insulator $(E_{fT}M_f)_{\theta}$ is the only state that has a simple mean field description on the pyrochlore lattice with Kramers spins!

Including fluctuations will lead to a lattice gauge theory with a $U(1)$ gauge field $a_{rr'} = -a_{r'r}$ described by the Hamiltonian 
\begin{equation}
H = -\sum_{\langle rr' \rangle} f^\dagger_r t_{rr'} e^{ia_{rr'}}f_{r'}
\end{equation}
(with the hopping matrix $t_{rr'} = t_1 + i t_2 \vec d_{rr'} \cdot \vec \sigma $) supplemented with the Gauss law constraint 
\begin{equation}
\sum_{r'}E_{rr'} = f^\dagger_r f_r - 1
\end{equation}
where $E_{rr'}$ is the integer electric field conjugate to $a_{rr'}$.  A guess for a spin Hamiltonian which may favor the topological Mott insulator state is obtained by performing a strong coupling expansion of this lattice gauge theory. The resulting Hamiltonian takes the same form as Eq.~\eqref{realh} for symmetry reason (after the appropriate standard rotation from the global basis of spin quantization axis to the local basis).  The relative magnitude of the coupling constants depend on the parameter $w=t_2/t_1$ as follows:

\begin{eqnarray}
\label{fit}
 J_{zz}/J&=&2w^2+8w-1, \nonumber \\
 J_{\pm}/J&=&2w^2+2w+\frac{1}{2}, \nonumber \\
  J_{\pm\pm}/J&=&w^2-2w+1, \nonumber \\
   J_{z\pm}/J&=&\sqrt{2}(-2w^2+w+1),
\end{eqnarray}
where $J$ is an overall constant, and the underlying fermion partons are gapped (and topological) when $w<-2$ or $w>0$ and $w\neq1$. A spin Hamiltonian was obtained in Ref.~\onlinecite{pesinlb} starting from a Hubbard model. The free fermion part of the Hubbard model in Ref.~\onlinecite{pesinlb}, in the limit of strong on-site spin-orbit coupling, corresponds to the same hopping Hamiltonian in Eq.~\eqref{mfhopping}, with $w=t_2/t_1=0.215$ determined by orbital physics. The spin Hamiltonian obtained in Ref.~\onlinecite{pesinlb}, after a basis rotation (see Ref.~\onlinecite{balentsqspice}), agrees with Eq.~\eqref{fit} with $w=0.215$. 
This simple consideration can provide a useful guide in searching for realizations of the topological Mott insulator in the family of rare earth pyrochlores if their exchange parameters can be determined by experiment. We should caution however that our arguments in this subsection are only suggestive, and a reliable determination of the phase diagram of spin models for these pyrochlores is currently beyond the reach of theoretical technology. On the experimental side, for $Yb_2Ti_2O_7$, a determination of the exchange parameters was provided in Ref. \onlinecite{rossetal}. However this has been disputed by newer experiments\cite{coldeakitp} which suggest instead a rather different set of parameters. In view of the existing  uncertainties in both the experiment and the theory we will leave further discussion of models and materials for the future.


\subsection{Experimental signatures}
\label{experimental}

Here we offer some suggestions on experiments that may help distinguish these different $U(1)$ spin liquids. 

First we ask about distinctions in neutron scattering experiments. Spin flip excitations that can scatter neutrons are created by local operators.  If the only global symmetry is time reversal, then neutrons will couple to all local operators that are odd under time reversal. Let us consider a few important ones. First since the emergent magnetic field is time reversal odd, neutrons can couple directly to the fluctuations of the internal magnetic field. As discussed in Ref.~\cite{gMFT} this enables neutron experiments to detect the emergent photon. Second the number density of magnetic monopoles, and the current of emergent electric charge are also local ${\cal T}$-odd 
operators.  Coupling to these will lead to an increase in the scattering cross-section when the energy transfer exceeds twice the gap of the $M$ and the $E$ particles. Generally there should be two thresholds in the scattering cross-section set by the $M$ and $E$ gaps.   If the $E$ particle is a Kramers doublet (which it is in some of the $7$ phases) then spin flip excitations can also form out of a pair of $E$ particles through a combination like $E^\dagger \vec \sigma E$. In this case the $E$- threshold may be much more sharply defined than in the case  where $E$ is a Kramers singlet.

Other useful information can be gleaned by studying the effects of  an applied magnetic field $\vec{B}$. Consider the phases where $E$ is a Kramers doublet and has a gap smaller than $M$ and other composite excitations. For a Kramers doublet $E$ particle, a direct coupling at quadratic level is allowed:
\be
\Delta H\sim T^{ij}B_iE^{\dagger}\sigma_jE,
\ee
where $T^{ij}$ is a tensor consistent with lattice symmetries. This implies that with increasing $\vec B$ the gap of the $E$ particle will close at some finite value before the $M$ gap closes. 
If $E$ is a boson, it will condense in such a $\vec{B}$ field and the $U(1)$ gauge field will be gapped. Since $E$ is a Kramers doublet, such a condensate necessarily breaks time-reversal symmetry. Therefore the resulting phase is magnetically ordered and can be probed through neutron scattering. 
If $E$ is a fermion, a fermi surface will emerge beyond the critical field, and the system becomes a $U(1)$ spin liquid with spinon fermi surface, which can be probed through heat capacity or heat transport measurements
(see Refs. \cite{ong,matsuda} for interesting recent heat transport measurements on pyrochlore magnets). 
Similar consideration also applies if the $M$ gap is smaller than the $E$ gap, and can indicate the statistics of the $M$ excitation in that case.  Thus the behavior in a magnetic field can provide useful information to partially distinguish these different $U(1)$ spin liquids. 

The two phases $(E_{fT}M_f)_{\theta}$ and $E_{bT}M_f$ necessarily have protected surface states. As we described it is likely that these surfaces are in gapless  phases  in which case it may be possible to detect the surface excitations.  A useful experiment will be to deposit a ferromagnet on the surface and measure the resulting thermal Hall effect. This kind of experiment  might perhaps be  interesting to explore  in $Yb_2Ti_2O_7$.

\section{Conclusion}
In this paper we have provided a detailed understanding of time reversal symmetric $U(1)$ quantum spin liquids. Our results were summarized in Sec. \ref{summ}. To conclude we highlight a few open questions. 
We have not discussed the effects of spatial symmetry at all. This will lead to a finer distinction between these spin liquid phases, and may impact the discussion of existing experimental candidates. A discussion of the effects of space group symmetry on time reversal symmetric $U(1)$ quantum spin liquids in a cubic lattice was provided in early work by Ref. \onlinecite{lesikts05} (see also subsequent related work in Ref. \onlinecite{bergman} on a model without time reversal on a pyrochlore lattice). 
Even with just time reversal it will be useful to identify sharp experimental fingerprints to distinguish the different phases. It may be interesting for future numerical work to study the loop wave functions described in Sec. \ref{loops} and explicitly demonstrate their correctness in describing the various spin liquids. 

{  Since the appearance of this paper on the arxiv, there have been a number of related further developments which we briefly summarize here. Our own subsequent work\cite{wstoappear} exploited the bulk duality of the topological Mott insulator described in this paper to provide a new `dual Dirac liquid' description of the surface of spin-orbit coupled electronic topological insulators.  This bulk duality, and the same dual Dirac liquid was independently obtained in Ref. \onlinecite{maxashvin} which appeared simultaneously with Ref. \onlinecite{wstoappear}.  See also the very recent paper\cite{maxtdf} with more results on the bulk duality. }

%




We thank Lucile Savary for useful discussions. This work was supported by NSF DMR-1305741.  This work was also partially supported by a Simons Investigator award from the
Simons Foundation to Senthil Todadri.

\appendix

\section{Model for $Yb_2Ti_2O_7$}
\label{qspicemdl}

Here we briefly review the spin Hamiltonian in Ref.~\cite{balentsqspice}, obtained by fitting the neutron scattering data in Ref.~\cite{rossetal} through spin wave theory.  The values of the parameters in this model have, however, been questioned in more recent work\cite{coldeakitp}. The Hamiltonian has only nearest-neighber terms and takes the form
 \begin{eqnarray}
 \label{realh}
    H=&\sum_{\langle ij\rangle}&J_{zz}S^z_iS^z_j-J_{\pm}\left(S_i^+S_j^-+S_i^-S_j^+\right) \nonumber \\
    &+&J_{\pm\pm}\left(\gamma_{ij}S_i^+S_j^+\gamma_{ij}^*S_i^-S_j^-  \right) \nonumber \\
    &+& J_{z\pm}\left[S^z_i\left(\zeta_{ij}S_j^++\zeta_{ij}^*S_j^- \right)+\left( i\leftrightarrow j\right) \right],
   \end{eqnarray}
where $S_i^{\mu}$ are spin coordinates in the local basis of spin ice. $\zeta_{ij}, \gamma_{ij}$ are $4\times4$ matrices acting within each tetrahedra and have the form
\be
\zeta=\lp\begin{array}{cccc}
          0 & -1 & e^{i\pi/3} & e^{-i\pi/3} \\
          -1 & 0 & e^{-i\pi/3} & e^{i\pi/3} \\
          e^{i\pi/3} & e^{-i\pi/3} & 0 & -1 \\
          e^{-i\pi/3} & e^{i\pi/3} & -1 & 0 \end{array}\rp, \gamma=-\zeta^*.
\ee

The coupling constants are, in $meV$,
\begin{eqnarray}
\label{realj}
 &J&_{zz}=0.17\pm0.04, \hspace{9pt} J_{\pm}=0.05\pm0.01, \nonumber \\
 &J&_{\pm\pm}=0.05\pm0.01, \hspace{7pt} J_{z\pm}=-0.14\pm0.01.
\end{eqnarray}
 
If $J_{zz}$ donimates over the other coupling constants, the system at low energy will be restricted to the spin ice manifold. But it is not clear whether the above Hamiltonian falls into such a regime, given that the other terms seem comparable to $J_{zz}$ in magnitude. Very recent experimental work\cite{coldeakitp} has disputed these parameter values - the revised values are substantially different and place the system even farther away from the regime where the restriction to the spin ice manifold is legitimate. 
   
\section{Time-reversal action on electric charge}
\label{tre}

In general, the fundamental electric particle $E$ can be multi-component with an internal index $i$. All the $E_i$ particles carry the same gauge charge, which means any object of the form $E^{\dagger}_iO_{ij}E_j$ must be gauge neutral and hence corresponding to a local operator.

In general, time-reversal could act on $E$ particles as
\be
\T E_i\T^{-1}=T_{ij}E_j,
\ee
where $T$ is a matrix. This implies that
\be
\T^2 E_i\T^{-2}=(T^*T)_{ij}E_j,
\ee
where the anti-unitarity of $\T$ was used. However, $\T^2$ on any local operator should be trivially identity. Therefore we should have
\be
(T^*T)^{\dagger}O(T^*T)=O,
\ee
for any matrix $O$. This can be true only if $T^*T=e^{i\phi}I$ where $I$ is the identity matrix. Combining with the complex conjugate relation $TT^*=e^{-i\phi}I$, we conclude that $e^{i\phi}=\pm1$. Therefore $\T^2 E\T^{-2}=\pm E$.

Notice that the above derivation assumes that all the local objects have $\T^2=1$ on them, which is true for a spin system. However, if charge-neutral Kramers fermion is present in the microscopic system (for example in a superconductor), then we do have local objects with $\T^2=-1$. Following the above logic, it will be possible to have $\T^2=\pm i$ for fractionalized objects such as $E$ particles.

\section{The same $U(1)$ spin liquid from gauging two distinct insulators}
\label{samegauging}

We start from fermions with $U(1)\times\T$ symmetry, which corresponds to the fermionic monopoles discussed in the main text. It is known that for free fermions the band structures of such kind of fermions are classified by an integer $n$. We now show that the $U(1)$ spin liquids obtained by gauging the fermionic monopoles with band topology $n$ is the same as that from gauging the fermions with band topology $-n$. The reason is very simple: at the level of band structures, the difference between a $+n$ band and a $-n$ band is the way time-reversal $\T$ is implemented. This can be most easily seen through the surface Dirac cones, where the effective Hamiltonian looks the same for bands at $\pm n$:
\be
H=\sum_{i=1}^n\psi^{\dagger}_i(p_x\sigma_x+p_y\sigma_z)\psi_i,
\ee
but time-reversal acts differently for $\pm n$: $\T\psi_i\T^{-1}=\pm i\sigma_y\psi^{\dagger}$.

Before gauging, the different time-reversal action makes it impossible to continuously tune one state into the other. But after gauging, the difference becomes simply a gauge transform $U=e^{i\pi Q}$ where $Q$ is the charge. Therefore the $\pm n$ bands become gauge equivalent, and give rise to the same quantum spin liquid.

It is also known\cite{3dfSPT2,maxvortex} that a band indexed by $n$ labels the same interacting bulk state as a band indexed by $n+8$. It is also known that bands indexed by $n$ and $n+4$ differ from a bosonic symmetry-protected topological state called $eTmT$ topological paramagnet\cite{avts12,hmodl}, which is protected only by time-reversal. Together with the previous identification of $n$ and $-n$ bands, we conclude that states with band index $n=\pm1\hspace{1pt}(mod\hspace{1pt}8)$ and $n=\pm3\hspace{1pt}(mod\hspace{1pt}8)$ corresponds to two distinct phases. The bulk excitations of these two phases are identical and correspond to the $(E_{fT}M_f)_{\theta}$ $U(1)$ spin liquid (topological paramagnet). The two phases have different surface states, and one can obtain one state from the other by combining with a $eTmT$ topological paramagnet.

As discussed in Sec.~\ref{addspt}, the $eTmT$ topological paramagnet becomes trivial when combined with the $E_{bT}M_f$ state. Therefore all the states with band index $n=2\hspace{1pt}(mod\hspace{1pt}4)$ corresponds to a unique $E_{bT}M_f$ state.

\section{Surface states of various $U(1)$ spin liquids}
\label{surfaceappendix}

Here we discuss the surface states of the first five phases in Table~\ref{u1gauge1}. All these phases have a particle ($E$ or $M$) that is a trivial boson. Therefore the ``wall'' corresponds to this particle is trivial. However, they can still have a nontrivial wall corresponding to the nontrivial quasiparticle. For example, the $E_bM_f$ phase could have an $M$-wall, with a $\Z_2$ topological order $\{1,e,m,\epsilon\}$ on the wall, where the $e$ and $m$ particles carry electric charge $q_E=1/2$. Following the logic in Ref.~\cite{metlitski}, this wall will convert the fermionic monopole into a bosonic one upon tunneling. The other $U(1)$ spin liquids have similar nontrivial walls, with a $\Z_2$ topological order and proper symmetry assignments on the $e$ and $m$ particles.

The physics of these ``walls'' will be important when we put two different spin liquids side by side. For example, if we put an $E_bM_b$ next to an $E_bM_f$, can we have coherent tunneling between the quasiparticles across the interface? Naively this is impossible since the monopole is fermionic in one region but bosonic in the other, and the system will just be two $U(1)$ spin liquids essentially decoupled from one another. However, we can put the $E$-wall described above on the interface, and the $M$ monopoles can now tunnel through the interface, with an $\epsilon$ particle left on the wall.

\section{Loop wavefunctions of various $U(1)$ spin liquids}
\label{wwloopappendix}

\begin{figure}
\begin{center}
\includegraphics[width=2in]{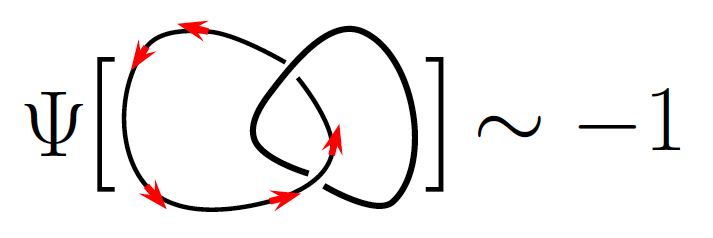}
\end{center}
\caption{Loop wavefunctions of the $E_{bT}M_b$, $E_{fT}M_b$ and other phases:  the amplitude changes sign whenever a directed loop links with an undirected one.  }
\label{wf6}
\end{figure}

Here we discuss loop wavefunctions of various other $U(1)$ spin liquids in the same spirit of Sec.~\ref{wwloop}. One can think of this approach as starting from a Walker-Wang wavefunction\cite{walkerwang,ashvinbcoh,fidkowski3d,chenanomaloussymm,fSTO3} and ``gauging'' the $U(1)$ symmetry. 
Many of these loop wavefunctions can be written in a form slightly different from that in Sec.~\ref{wwloop}. The wavefunctions have a directed ``half'' loop and an undirected loop condensing simultaneously, with a $(-1)$ phase in the wavefunction whenever a ``half'' loop and an undirected loop mutually link (see Fig.~\ref{wf6}). Following similar analysis in Sec.~\ref{wwloop}, the resulting phase is a $U(1)$ spin liquid, with two fundamental particles $E$ and $M$. One of them is bound with the open end of an undirected loop, and the other is the open end of a doubled half loop. 

If the directed loops reverse their directions under time-reversal, they are magnetic loops; otherwise they are electric loops.
It is now easy to see how to assign various quantum numbers to the dual particles in this description. For example, in the magnetic loop picture, we can put a Haldane chain in the undirected loop to make the $E$ particle Kramers, in which case we obtain the $E_{bT}M_b$ phase in the magnetic loop picture. If we also make the undirected loop a ribbon with a $(-1)$ self-linking sign, we convert the $E$ particle to a fermion and obtain the $E_{fT}M_b$ phase in the magnetic loop picture.

There are some more complicated cases, including the $E_{bT}M_f$ phase in the magnetic loop picture, and the $(E_{fT}M_f)_{\theta}$ in both pictures. The surface topological order of these phases are well studied in the literature, from which one can derive the corresponding Walker-Wang wavefunctions and their gauged version\cite{fidkowski3d,fSTO1,fSTO2,fSTO3,fSTO4,3dfSPT2,maxvortex}. However, the results are somewhat complicated (especially for $(E_{fT}M_f)_{\theta}$ which is non-Abelian) and not particularly illuminating, so we will obmit the discussion here.

\end{document}